# MEC Task Offloading in AIoT: A User-Centric DRL Model Splitting Inference Scheme

LI Weixi[1], GUO Rongzuo[1*], WANG Yuning [2], CHEN Fangying[1]

1. College of Computer Science, Sichuan Normal University, Chengdu 610101, China;

2. Academic Affairs Office, Sichuan Water Conservancy Vocational College, Chongzhou 611231, China.

**Abstract: With the rapid development of the Artificial Intelligence of Things (AIoT), mobile edge computing (MEC) has become an essential technology underpinning AIoT applications. However, multi-angle resource constraints, multi-user task competition, and the complexity of task offloading in dynamic MEC environments pose new technical challenges. To address these, we propose a user-centric deep reinforcement learning (DRL) model splitting (UCMS) inference scheme. This scheme combines a user-server co-selection algorithm with a UCMS_MADDPG-based offloading algorithm to realize efficient inference responses in dynamic environments with multi-angle resource constraints. Specifically, we formulate a joint optimization model that integrates resource allocation, server selection, and task offloading, aiming to minimize the weighted sum of task delay and energy consumption. After decoupling the optimization, the user-server association is handled through a co-selection algorithm. To address the mixed decision problem, we design an algorithm centered on user pre-decision that splits the action space into user-side and server-side components to coordinate continuous and discrete decision outputs. In addition, a priority sampling mechanism based on a reward-error trade-off is introduced to enhance experience replay. Simulation results show that the proposed UCMS_MADDPG-based offloading algorithm demonstrates superior overall performance compared with**

**other benchmark algorithms in dynamic environments.**

This work was supported by Natural Science Foundation of China (11905153). Weixi Li, female, master, research direction is mobile edge computing, E-Mail: liweixi@stu.sicnu.edu.cn; Rongzuo Guo, male, master, professor, master's supervisor, research directions include embedded systems, IoT perception technology, system reliability, and intelligent control, E-Mail: gyz00001@163.com.

# 1 Introduction

The convergence of artificial intelligence (AI) and Internet of Things (IoT) has resulted in the Artificial Intelligence of Things (AIoT)[1], which holds broad application potential across diverse fields. By integrating the automated analysis and decision-making capabilities of AI with the extensive connectivity and real-time data acquisition of IoT, AIoT enhances system intelligence and facilitates the digital transformation of traditional industries. However, AIoT faces significant technical challenges in practical applications, primarily due to the massive volume and heterogeneity of data generated by IoT devices, which place heavy demands on computing architectures. Traditional cloud computing (CC) relies on centralized data centers for data processing, storage, and analysis, and has previously shown considerable adaptability to IoT applications[2]. Nevertheless, with the rapid growth of IoT devices and the diversification of application scenarios, traditional cloud computing architectures struggle to meet the stringent demands of AIoT for real-time responsiveness, efficiency, and sustainable energy use. This limitation results from high transmission latency, inefficient bandwidth utilization, and substantial energy consumption associated with centralized processing[3]. To address these issues, mobile edge computing (MEC) has emerged as a promising solution[4]. By offloading computing, storage, and networking functions to the network edge near end devices, MEC enables a computing paradigm characterized by low latency, improved bandwidth utilization, and enhanced energy efficiency. Despite its advantages in improving the computational efficiency of AIoT systems, MEC continues to face challenges such as limited computing capacity and restricted resources. Therefore, intelligent resource management and optimized task offloading strategies remain critical research problems in MEC.

The task offloading of MEC in AIoT is commonly addressed using deep reinforcement learning (DRL), where the computational task offloading process is modeled as a Markov decision process (MDP) to optimize decision-making within the MEC system[5-8]. For instance, Wu et al.[9] proposed a hybrid offloading strategy that integrated convex optimization with a deep Q-network (DQN) to optimize task offloading in time-varying fading channels, demonstrating strong performance in large-scale networks. Hu et al.[10] addressed task offloading under time-varying channels by formulating it as minimizing the average long-term service cost, jointly considering power consumption and buffer delay in dynamic task patterns. To tackle the mixed action space involving continuous and discrete decisions, they employed a combination of deep deterministic policy gradient (DDPG) and dueling double deep Q-network (D3QN). Compared with DQN, the proposed algorithm achieved significant performance gains. Due to limited computing capacity and resources of user devices and MEC servers, task offloading algorithms must effectively handle discrete, continuous, and hybrid action spaces. However, both DQN and DDPG exhibit limitations when applied to large discrete action spaces or hybrid action spaces that combine continuous and discrete decisions[11, 12], restricting their practicality in complex real-world environments. Although task offloading algorithms based on multi-agent deep deterministic policy gradient (MADDPG) are well-suited to continuous action spaces[13], effectively representing and jointly optimizing both discrete and continuous action spaces remains a major challenge[14].

Numerous existing studies have considered the constraints imposed by limited wireless communication resources and server-side computational capacity, including the restricted local computing power and finite battery energy of user devices[15, 16], and computing resource constraints on the MEC server[17]. To address these challenges, various solutions have been proposed, such as exploiting cloud-edge collaboration (MEC-MCC)[18] and enabling cooperative computation among multiple MEC servers[19]. However, server storage constraints have received limited attention in existing DRL-based task offloading algorithms[20-26].

Specifically, this paper mainly considers the key technical challenges of MEC systems in current AIoT scenarios. It focuses on the joint optimization of multi-angle resources, including communication, computation, and storage, under dynamic conditions; the effective management of hybrid action spaces covering continuous resource allocation and discrete offloading decisions; the coordination of multiple users and multiple servers under competing tasks in MEC systems; and the limitations of existing DRL methods in dealing with these complex problems simultaneously. These challenges are particularly pronounced in practical deployments, where server storage constraints significantly impact system performance, an aspect often overlooked in prior research.

To better align with real-world environments, we simulate the multi-angle resource constraints on user devices and MEC servers, and adopt a user-centric model splitting inference scheme to enable task offloading decision-making and response. It is worth clarifying that we use “model splitting inference” to specifically refer to the splitting of the decision-making process between users and servers in our user-centric model splitting inference framework in this paper. That is, the user-side actor network generates a preliminary offloading decision, while the edge server then approves or rejects the request based on global resource information. This definition is distinct from conventional deep neural network (DNN) layer splitting[27] and better reflects the hierarchical decision mechanism considered in this work. With a suitable splitting strategy, this approach can significantly reduce inference delay and end-side energy consumption, at the cost of some communication overhead[28].

The key contributions of this paper include:

- We propose a dynamic MEC environment with multiple users and servers in AIoT, focusing on the task processing problem of end users in service areas where multiple servers overlap. The model considers multi-angle resource constraints, including communication resources, user device resources, user computing capacity, and server storage capacity.
- To simplify the optimization problem, we decouple it into a user-server selection problem and a task offloading problem. We design a user-server co-selection algorithm to address the selection and matching between users and servers.
- A hybrid decision support UCMS_MADDPG algorithm, using a user-centric model splitting inference scheme. With user decisions as the primary driver, servers contribute to offloading decisions. The pre-decision for resource allocation and task offloading is first made on the user-side CPU (first-stage model inference) and then refined on the server-side CPU to complete the hybrid decision process (second-stage model inference).

- We perform comprehensive experiments, including ablation analysis and comparisons with various heuristic baselines, to verify the proposed algorithm's performance.

The structure of this paper is as follows. Section 2 reviews the related work. In Section 3, we present the system model and formulate the optimization problem. Section 4 analyzes the problem and proposes corresponding solutions. Section 5 discusses the simulation results. Finally, Section 6 concludes the paper.

# 2 Related Work

Many scholars have conducted extensive research on task offloading strategy optimization in MEC. Among these approaches, DRL techniques for task offloading and resource allocation in MEC environments have emerged as a primary focus. Li et al.[6] focused on joint computation and communication optimization for user devices handling multiple tasks in unmanned aerial vehicle (UAV)-assisted MEC networks, introducing a multi-agent proximal policy optimization (MAPPO) framework with beta distribution to jointly optimize UAV trajectories, task partitioning, and overall weighted energy consumption. Yan et al.[7] applied a long short-term memory (LSTM) network enhanced with an attention mechanism, integrated with the DDPG algorithm, to reduce system latency in vehicular MEC networks. Zhao et al.[29] investigated the joint optimization problem of computation offloading, resource allocation, and charging scheduling, and proposed a task offloading scheme based on the twin delayed deep deterministic policy gradient (TD3) algorithm. By leveraging the delayed policy update and clipped double Q-learning mechanisms, their approach effectively improves training stability in complex continuous action spaces. Mi et al.[30] developed a multi-agent online control (MAOC) algorithm to handle delay-sensitive tasks in device-to-device (D2D)-assisted MEC systems, adjusting CPU frequencies in discrete time slots to make intelligent decisions, thereby reducing edge server load and preventing network congestion. Hu et al.[31] developed a multi-agent deep reinforcement learning (MADRL) algorithm that integrates an actor-critic (AC) framework, embedding techniques, and the centralized training and decentralized execution (CTDE) paradigm. The proposed method successfully reduces the long-term system energy consumption in delay-sensitive cellular networks with multiple user tasks. The above research demonstrates the effectiveness of DRL in resource allocation and task offloading optimization.

However, most of these works focus on a single performance metric, such as system latency or energy consumption, which limits their ability to comprehensively characterize system behavior in dynamic environments. In addition, solving the mixed problem of continuous and discrete action spaces has become a major research focus in DRL approaches.

In dynamic MEC environments, many scholars focus on multi-objective joint optimization, addressing communication and computing resource constraints to enhance system performance. This remains a complex challenge. Naqqash Tariq et al.[17] proposed a DRL-driven energy-efficient task offloading (DEETO) scheme that considers battery capacity and computing power constraints in UAV emergency scenarios. By implementing a hybrid task offloading mechanism, DEETO minimizes UAV energy consumption. Nguyen et al.[21] introduced a collaborative task offloading and block mining (TOBM) framework for blockchain-enabled MEC systems, aiming to maximize overall system utility. Dong et al.[22] decomposed the problem into two parts: they used the TD3 algorithm to solve resource allocation (RA) under fixed task offloading decisions, and applied heuristic particle swarm optimization (PSO) to optimize task offloading guided by optimal value functions. Their approach reduces production time and energy consumption in industrial IoT scenarios. Gong et al.[32] examined dependent task adaptive offloading in dynamic networks and proposed a DRL-based dependent task offloading strategy (DTOS) to reduce latency and energy consumption in network services. Zhang et al.[33] proposed a multi-head self-attention enhanced multi-agent deep dirichlet deterministic policy gradient (MHSA-MAD3PG) algorithm to address the joint optimization of MEC task offloading and resource allocation in scenarios involving multiple mobile devices. Their approach enables the coordinated optimization of system utility, service latency, and energy consumption.

To address diverse resource constraints, several studies have explored MEC-MCC techniques and collaborative computation across multiple servers. Zhang et al.[15] investigated highly dynamic MEC systems with energy harvesting devices and proposed a multi-device (MD) hybrid decision AC algorithm for dynamic task offloading. This algorithm manages the hybrid action space, including both continuous and discrete decisions across devices, thereby balancing latency and energy consumption. Zhao et al.[16] proposed a DRL algorithm with a hierarchical reward function for video offloading in MEC networks. This approach ensures video security, improves user quality of experience (QoE), and reduces energy consumption. Wu et al.[18] developed an energy-efficient dynamic task offloading (EEDTO) algorithm for MEC-MCC environments supporting blockchain, optimizing the trade-off among computing power constraints, high latency, secure offloading, and data integrity. Yang et al.[19] introduced a global state-sharing model based on a variational recurrent neural network (VRNN) in distributed MEC systems, reducing server communication overhead and task execution time. Du et al.[23] addressed real-time, efficient processing of intensive applications in healthcare IoT devices. They proposed a collaborative cloud-edge offloading model tailored for ultra-dense edge computing (UDEC) networks, reducing both energy consumption and task execution latency. Zakaryia et al.[24] proposed a cooperative multi-UAV-assisted MEC framework, in which the positions of unmanned aerial vehicles, cooperative task offloading decisions, and computational and communication resource allocation are jointly optimized to maximize the long-term system resource utilization. In real-world environments, edge server resources are often constrained. Although the above studies consider the computational resource constraints of servers, they largely overlook the constraints of storage resources. Setting server storage resources to an ideal state in task-intensive scenarios with many users may improve system utility, but this assumption reduces applicability to realistic settings.

Table 1 summarizes a comparative analysis of representative existing works. A critical analysis of Table 1 reveals that while DRL has proven effective in addressing complex joint optimization problems in MEC systems and has made significant progress in MEC task offloading, several key limitations remain. First, only a limited number of approaches effectively handle the hybrid action space problem that combines continuous and discrete decisions. Second, many algorithms are evaluated in relatively static environments without fully addressing the challenges of dynamic user-server associations and time-varying tasks. Third, most research on dynamic environments gives limited consideration to the storage resource constraints of servers, assuming ideal storage conditions that do not reflect real-world deployments. These limitations directly motivate the comprehensive approach proposed in this work.

As highlighted in the above review, our approach provides a comprehensive solution that simultaneously addresses multiple limitations in the analysis. In the context of a dynamic MEC environment with multiple users and servers, we consider the multi-angle resource constraints of user devices and servers. The objective is to jointly optimize resource allocation, server selection, and task offloading strategies, while minimizing the weighted sum of task execution latency and energy consumption, thereby improving the quality of service.

**Table 1 Comparison of Existing Jobs**

| Paper | Objective | Method | Multiple servers | Dynamic environment | Energy constraint | Server computing resource constraint | Server storage resource constraint |
|---|---|---|---|---|---|---|---|
| Li et al.[6] | Minimize energy consumption | b-MAPPO | √ | √ | × | √ | × |
| Yan et al.[7] | Minimize latency | LDDPG | × | × | √ | × | × |
| Zhao et al.[29] | Maximize system utility | TD3 | √ | √ | √ | √ | × |
| Mi et al.[30] | Minimize latency | MAOC | × | √ | √ | × | × |
| Hu et al.[31] | Minimize energy consumption | CTDE-MAAC | √ | √ | × | √ | × |
| Naqqash Tariq et al.[17] | Minimize latency and energy consumption | DEETO | × | √ | √ | √ | × |
| Nguyen et al.[21] | Maximize system utility | MADDPG | × | √ | × | × | × |
| Dong et al.[22] | Minimize latency and energy consumption | TD3 and PSO | × | √ | × | × | × |
| Gong et al.[32] | Minimize latency and energy consumption | DTOS | × | √ | × | × | × |
| Zhang et al.[33] | Minimize latency and energy consumption | MHSA-MAD3PG | √ | √ | √ | √ | × |
| Zhang et al.[15] | Minimize latency and energy consumption | MD Hybrid AC | √ | × | √ | √ | × |
| Zhao et al.[16] | Minimize energy consumption and maximize QoE | JVFRS-CO-RA-MADDPG | √ | √ | √ | √ | × |
| Wu et al.[18] | Minimize latency and energy consumption | EEDTO and Lyapunov | √ | √ | × | × | × |
| Yang et al.[19] | Minimize latency | VRNN | √ | √ | × | √ | × |
| Du et al.[23] | Minimize latency and energy consumption | DDPG | √ | √ | × | × | × |
| Zakaryia et al.[24] | Maximize system utility | DDPG-DTLCM | √ | √ | × | √ | × |

# 3 System Model

The multi-edge server deployment in the MEC environment for AIoT is illustrated in Fig. 1. In this model, we assume that servers do not have direct physical links for communication and independently process computing tasks within their respective service coverage areas. However, due to overlapping service regions between servers, an end user located in such a region can select to offload tasks to a specific server.

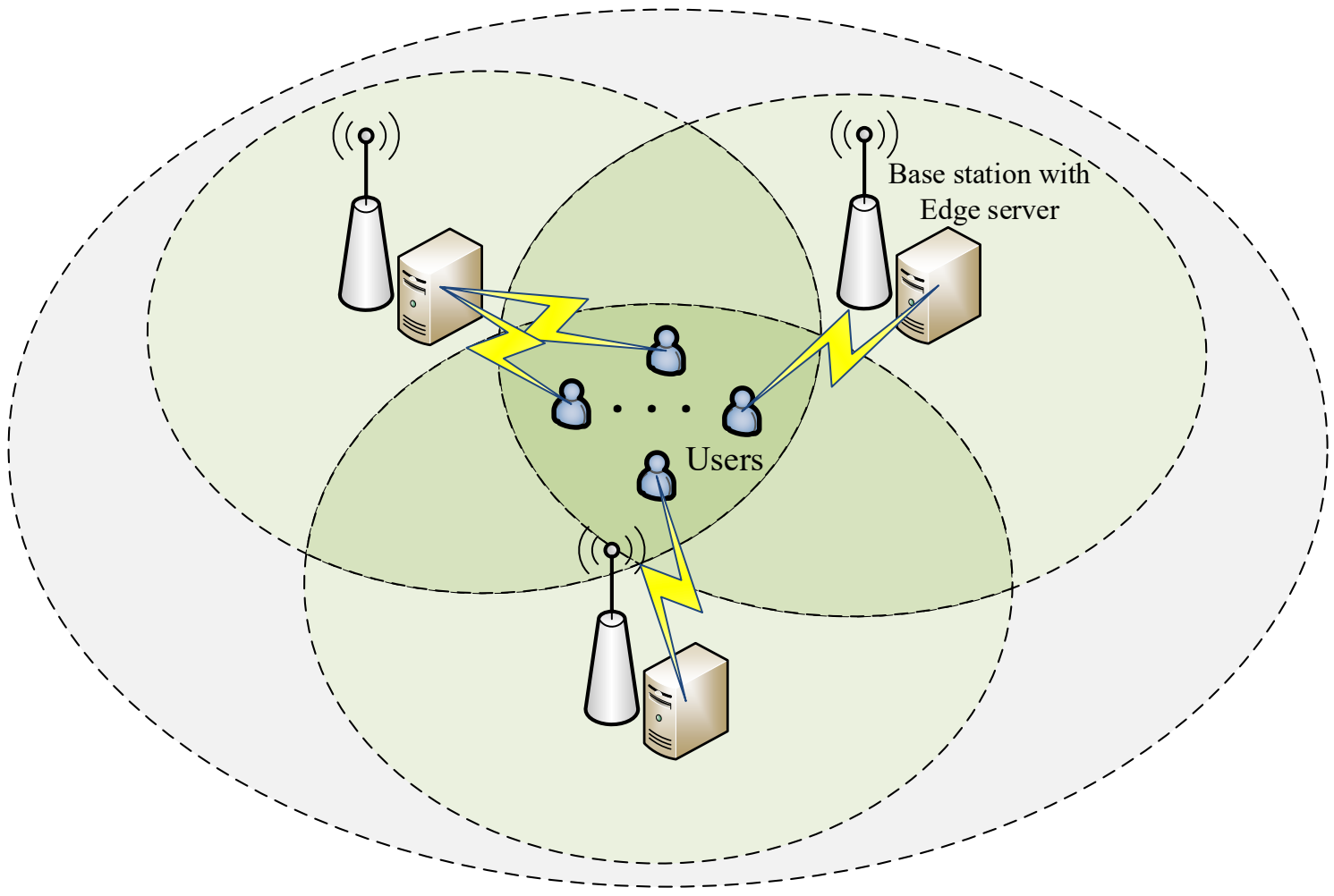

**Fig.1 System environment layout**

This section introduces the task model of MEC system in AIoT scenarios, as shown in Fig. 2. We consider a set of $M$ edge servers (ESs) with overlapping service areas and $N$ user devices (UDs), each carrying out different tasks. The set of edge servers is denoted as $M = \{1, 2, \ldots, m\}$, and the set of user devices is denoted as $N = \{1, 2, \ldots, n\}$. Each ES is equipped with multiple CPUs and a fixed storage capacity to provide both computing and storage services, while each UD is equipped with a single CPU and battery. Task data are transferred between ESs and UDs through a base station (BS) over a wireless transmission channel. The following section provides a detailed introduction to the tasks, computing, and energy harvesting models.

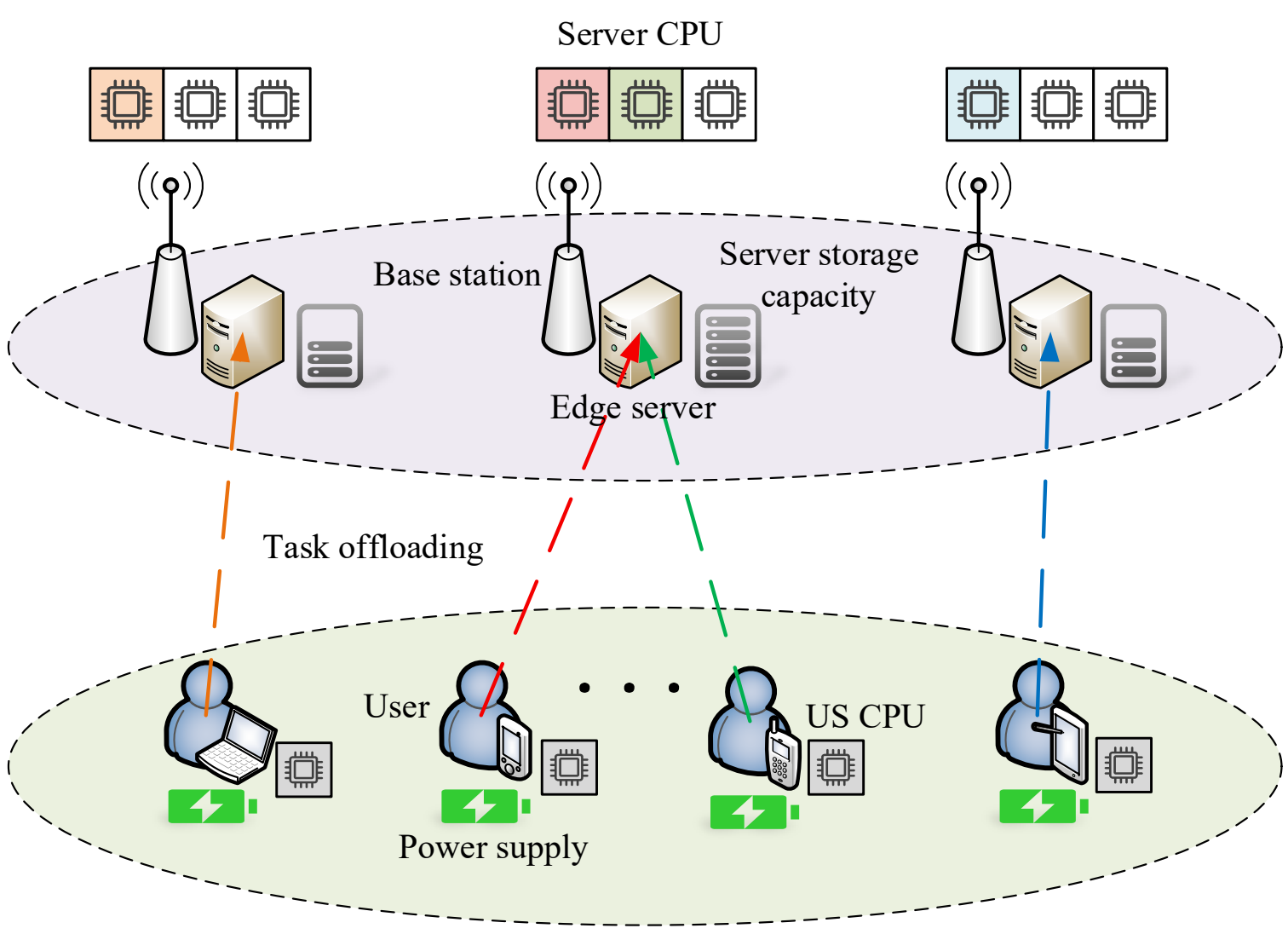

**Fig.2 System task model**

## 3.1 Task Model

Usually, the UD has limited computing power, and computing tasks are sensitive to delay. Therefore, we model task execution within discrete time slots, where $\varphi_{max}(t)$ denotes the maximum tolerated delay for task execution in each slot *t*, and the set of operational periods is denoted as $T=\{1,2,\ldots,t\}$. Each user device UD$n \in N$ is assumed to generate a random task at the beginning of each time slot $t \in T$, and if the task is not completed within $\varphi_{max}(t)$, it will be discarded before the next time slot begins. The task generated by UD$n$ in time slot $t$[1] is denoted as $(\beta_n(t), c_n(t), \varphi_n(t))$, where $\beta_n(t)$ represents the task size in bytes, and $c_n(t)$ denotes the number of CPU cycles required to process the task. The maximum allowable completion time for the task is given by $\varphi_n(t)$.

In an AIoT edge computing environment, the UD typically utilizes local computing, full offloading computing, or partial offloading computing to process tasks. In this paper, we focus on binary decision methods, specifically considering local computing and full offloading computing for handling UD tasks. The binary decision method $\boldsymbol{x}=\{x_n(t) \mid n \in N, t \in T\}$ is defined in Equation (1):

$$x_n(t)=\begin{cases}1, & \text{MEC computing} \\ 0, & \text{Local computing}\end{cases} \tag{1}$$

if the UD$n$ selects local computing, then $x_n(t)=0$, and if the UD$n$ selects offloading computing, then $x_n(t)=1$.

For a UD$n$ in overlapping service regions, connections can be established with multiple ES, but only one ES can be selected by UD for task offloading. Similarly, a binary decision is used to represent the UD selection, which $\boldsymbol{y}=\{y_{n,m}(t) \mid n \in N, m \in M, t \in T\}$ is defined in Equation (2):

$$y_{n,m}(t)=\begin{cases}1, & \text{selection} \\ 0, & \text{otherwise}\end{cases} \tag{2}$$

[1] All the following formulas are evaluated within a single time slot *t*.

if the UD$n$ selects the ES$m \in M$ for computing during offloading, then $y_{n,m}(t)=1$, otherwise $y_{n,m}(t)=0$.

Denote $Z_m = \{z_m(t) \mid m \in M, t \in T\}$ as the set of UD$n$ successfully connected to the ES$m$, where $z_m(t) = \sum_{n \in N} 1_{y_{n,m}(t)=1}$. To be more realistic, we stipulate that the UD connections to each ES are limited, and define the maximum user capacity of the ES$m$ as $z_{max}$, i.e., $z_m(t) \le z_{max}$.

## 3.2 Local Computing Model

In the local computing model, each UD$n$ is allocated a minimum and maximum local computing capacity, denoted as $f_{min}^l$ and $f_{max}^l$ (in GHz), respectively. This paper adopts a user-centric task offloading model, where if the UD$n$ decides to compute the task locally, the ES$m$ does not intervene. Furthermore, if the UD$n$ requests task offloading to the ES$m$, but the ES$m$ does not make a decision and opts to offload the task, the task of UD$n$ will still be processed locally. Details are described in Section 4.2.

Specifically, the local computing task delay $\tau_n^l(t)$ is determined according to the task size of the UD$n$ and the allocation of computing capacity:

$$\tau_n^l(t) = \frac{\beta_n(t)c_n(t)}{f_n^l(t)} \tag{3}$$

where $f_n^l(t)$ is the computing power of UD$n$ in time slot $t$, and the local computing power of UD$n$ is limited to its allocation budget, i.e., $f_{min}^l \le f_n^l(t) \le f_{max}^l$.

The energy consumption of the UD$n$ equipped with a CPU in each computing cycle can be calculated by $\kappa\left(f_n^l(t)\right)^2$. Similarly, $e_n^l(t)$ represents the energy consumption of local computing tasks:

$$e_n^l(t) = \kappa\beta_n(t)c_n(t)\left(f_n^l(t)\right)^2 \tag{4}$$

where $\kappa$ is the effective energy consumption factor depending on the UD$n$ chip architecture.

## 3.3 Offloading Computing Model

For the offloading computing model, to successfully offload its task to an ES, we specify that the minimum and maximum transmission power allocation budgets for each UD$n$ are expressed as $p_{min}^{t}$ and $p_{max}^{t}$ (in dBm), respectively. The number of CPUs equipped with each ES$m$ is $U_m$, the storage capacity constraint is denoted as $D_m$, and the computing power per CPU is denoted as $f_m^s$ (in GHz). At the same time, we consider that the wireless transmission channel follows a fading model, where the channel varies across different time slots but remains constant within the same time slot $t$. In the user-centric task offloading mode, if a UD$n$ makes an offloading request and the ES$m$ selected by it accepts the task, the task on the UD$n$ will be passed to the ES$m$, and the ES$m$ will select a CPU in $U_m$ for task computing. Details are also described in Section 4.2.

Similar to the approach in[21], we design the wireless network bandwidth W (in MHz) to be evenly allocated among K subchannels. For simplicity, the model in this paper makes several assumptions: each wireless transmission subchannel is used exclusively by one UD's tasks at a time, and interference between UDs is not considered. Additionally, the channel gain between the UD$n$ and the ES$m$ is denoted as $h_{n,m}(t)$, and constant noise power $\sigma^2$ is assumed, with the noise level remaining fixed at a given distance. The normalized channel gain for the uplink channel is then defined as:

$$g_{n,m}(t) = \frac{h_{n,m}(t)}{\sigma^2} \tag{5}$$

The transmission of tasks from the UD$n$ to ES$m$ requires transmission power, denoted as $p_{n,m}^{t}(t)$, which is constrained by the transmission power budget of UD$n$, i.e., $p_{min}^{t} \le p_{n,m}^{t}(t) \le p_{max}^{t}$. According to Shannon's capacity theorem, the transmission rate of a single transmission channel in the wireless network is given by $v_{n,m}(t)$, as shown in equation (6):

$$v_{n,m}(t) = \frac{W}{K} \log_2 \left(1 + p_{n,m}^{t}(t) g_{n,m}(t)\right) \tag{6}$$

Compared to the task data offloaded from the UD$n$ to ES$m$, the data size of computation results returned by the ES$m$ is much smaller. Therefore, similar to many other task offloading studies[15, 21], we assume that the transmission delay for returning the computation results from the ES$m$ to UD$n$ is negligible. Consequently, the transmission delay for offloading the computing task is denoted as $\tau_{n,m}^{o}(t)$, and it is determined by the transmission rate as follows:

$$\tau_{n,m}^{o}(t)=\frac{\beta_n(t)}{v_{n,m}(t)} \tag{7}$$

After the UD$n$ task is offloaded to ES$m$, the ES$m$ must process the offloading tasks of all UDs in each time slot. Since the tasks of UD$n$ have deadline constraints, the execution delay of computing tasks on ES$m$ must be carefully considered. Given that $U_m$ is limited and each CPU on ES$m$ can handle only one task at a time, the tasks are processed in order when they arrive, which is determined by the transmission delay $\tau_{n,m}^{o}(t)$. Based on the arrival order, tasks are assigned to available CPUs in sequence. If no CPU is available when a task arrives, the task must wait until a CPU becomes free. Therefore, we divide the task execution delay on ES$m$ into two parts: the queuing delay while waiting for the earliest available CPU, and the processing delay of the CPU for task computation.

The queuing delay, denoted as $\tau_{n,m}^{q}(t)$, represents the estimated time before the earliest executable CPU on ES$m$ can begin processing the UD$n$ task. This delay is determined by the completion times of previously accepted offloading tasks on ES$m$ that arrived before the UD$n$ task. This approach is adapted from the work of[34]. The queuing delay is then given by:

$$\tau_{n,m}^{q}(t)=\max\left(0,\min_{k=1}^{U_m}F_{m,k}(t)-\tau_{n,m}^{o}(t)\right) \tag{8}$$

where the term $\min_{k=1}^{U_m}F_{m,k}(t)$ represents the earliest time when any CPU on ES$m$ becomes available.

The processing latency, denoted as $\tau_{n,m}^{p}(t)$, is determined by the size of the UDn task and the CPU processing power of ES$m$:

$$\tau_{n,m}^{p}(t)=\frac{\beta_n(t)c_n(t)}{f_m^{s}(t)} \tag{9}$$

Therefore, the total delay of task offloading from the UD$n$ to ES$m$ is expressed as $\tau_{n,m}^{a}(t)$, which is determined by transmission delay, queuing delay, and processing delay:

$$\tau_{n,m}^{a}(t)=\tau_{n,m}^{p}(t)+\max\left(\tau_{n,m}^{o}(t),\tau_{n,m}^{q}(t)\right) \tag{10}$$

Since the ES in the model is directly powered by the power grid, the energy consumption for the execution of tasks on ES$m$ is not considered. Instead, we focus only on the energy consumption incurred by UD$n$ during the offloading process to ES$m$. The energy consumption for offloading a computing task is denoted as $e^{o}_{n,m}(t)$, and is defined as follows:

$$e^{o}_{n,m}(t) = p^{t}_{n,m}(t)\tau^{o}_{n,m}(t) \tag{11}$$

According to the two task calculation models of local and offloading described in Sections 3.2 and 3.3, the total delay and total energy consumption of the UD$n$ in binary decision mode can be obtained, denoted as $\tau_n(t)$ and $e_n(t)$:

$$\tau_n(t) = \left(1 - x_n(t)\right)\tau^{l}_{n}(t) + x_n(t)\tau^{a}_{n,m}(t) \tag{12}$$

$$e_n(t) = \left(1 - x_n(t)\right)e^{l}_{n}(t) + x_n(t)e^{o}_{n,m}(t) \tag{13}$$

## 3.4 Energy Harvesting Model

For the energy harvesting model, we define the minimum and maximum battery thresholds for each UD$n$ as $b^{h}_{min}$ and $b^{h}_{max}$ (in MJ), respectively. The battery energy is denoted as $b_n(t)$, which is constrained by its power threshold. Energy harvesting work adapted from[15]. Each UD$n$ is assumed to start fully charged at the beginning of the first time slot, and from then on, energy is harvested at the beginning of each subsequent time slot. The harvested energy is random, denoted as $d_n(t)$. The battery energy is primarily influenced by the energy consumed through local computation and offloading transmissions. Consequently, the battery energy available in the next time slot depends on both the energy consumed and the energy harvested during the current time slot. The battery energy update rule is governed by the following equation:

$$b_n(t+1) = \min\left(\max\left(b_n(t) - e_n(t) + d_n(t), 0\right), b^{h}_{max}\right) \tag{14}$$

The battery energy update rule ensures that the battery level of UD$n$ remains within the valid range, meaning it does not drop below zero or exceed the maximum power threshold. For convenience, Table 2 summarizes the key symbols used in the proposed system model.

**Table 2 Description of Key Notations Definitions**

| Notation | Definition |
| --- | --- |
| $M$ | Number of edge servers |
| $N$ | Number of user devices |
| $T$ | Period of operation |
| $\varphi_{max}(t)$ | Maximum tolerated delay |
| $\beta_n(t)$ | Task byte size of the UD*n* at time slot *t* |
| $c_n(t)$ | The number of CPU cycles required by the UD*n* to process the task |
| $\varphi_n(t)$ | Maximum time deadline for the expected completion of a UD*n* task |
| $x_n(t)$ | Task offloading decision of the UD*n* |
| $y_{n,m}(t)$ | Selection decision from the UD*n* and the ES*m* |
| $\tau_n^l(t)$ | Local computing delay of the UD*n* at time slot *t* |
| $f_n^l(t)$ | Local computing power of the UD*n* at time slot *t* |
| $e_n^l(t)$ | Local computing energy consumption of the UD*n* at time slot *t* |
| $z_{max}$ | Maximum user capacity of the ES*m* |
| $U_m$ | The number of CPUs the ESm is equipped with |
| $D_m$ | Maximum storage capacity of the ES*m* |
| $f_m^s$ | The CPU power of the ES*m* |
| $g_{n,m}(t)$ | Normalized channel gain from the UD*n* to the ES*m* |
| $p_{n,m}^t(t)$ | Transmission power of the UD*n* at time slot *t* |
| $v_{n,m}(t)$ | Channel transmission rate from the UD*n* to the ES*m* |
| $\tau_{n,m}^o(t)$ | Offloading computing delay of the UD*n* at time slot *t* |
| $e_{n,m}^o(t)$ | Offloading computing energy consumption of the UD*n* at time slot *t* |
| $b_n(t)$ | Battery energy of the UD*n* at time slot *t* |
| $d_n(t)$ | Energy harvesting by the UD*n* at time slot *t* |

## 3.5 Optimization Problem Modeling

According to the system model established in the previous text, each UD must decide whether to offload its tasks to an ES, considering the resource constraints of both the UD and the ES. The goal is to minimize the UD's computational cost by reducing both computation delay and energy consumption over the long term. To achieve this, we jointly consider total delay and energy consumption incurred by the UD in a binary decision-making model. We introduce weight coefficients $\rho_1$ and $\rho_2$, which are used to calculate the cost function of the task:

$$C_n(t)=\rho_1\tau_n(t)+\rho_2 e_n(t) \tag{15}$$

To minimize the average offloading cost across all UDs and run cycles *T*, we jointly optimize the ES selection decision, task offloading decision, local computing power allocation, and transmission power budget allocation, represented by $\boldsymbol{y}$, $\boldsymbol{x}$, $\boldsymbol{f}=\{f_n^l(t)\,|\,n\in N,t\in T\}$, and $\boldsymbol{p}=\{p_{n,m}^t(t)\,|\,n\in N,m\in M,t\in T\}$. In summary, the optimization problem P1 is given as follows:

$$\text{P1:}\quad \underset{\boldsymbol{y},\boldsymbol{x},\boldsymbol{p},\boldsymbol{f}}{minimize}\ \frac{1}{T}\sum_{t=1}^{T}\sum_{n=1}^{N}C_n(t) \tag{16a}$$

$$s.t.\quad \sum_{m\in M,n\in N} y_{n,m}(t)=1,\qquad \forall n\in N,\forall m\in M,\forall t\in T \tag{16b}$$

$$x_n(t)\in\{0,1\},\qquad \forall n\in N,\forall t\in T \tag{16c}$$

$$p_{min}^t\le p_{n,m}^t(t)\le p_{max}^t,\quad \forall n\in N,\forall m\in M,\forall t\in T \tag{16d}$$

$$f_{min}^l\le f_n^l(t)\le f_{max}^l,\qquad \forall n\in N,\forall t\in T \tag{16e}$$

$$b_{min}^h\le b_n^h(t)\le b_{max}^h,\qquad \forall n\in N,\forall t\in T \tag{16f}$$

$$\varphi_n(t)\le\varphi_{max}(t),\qquad \forall n\in N,\forall t\in T \tag{16g}$$

$$z_m(t)\le z_{max},\qquad \forall m\in M,\forall t\in T \tag{16h}$$

$$\sum_{n\in Z_m} x_n(t)\le K,\qquad \forall n\in N,\forall m\in M,\forall t\in T \tag{16i}$$

$$\sum_{n\in Z_m} x_n(t)\beta_n(t)\le D_m,\quad \forall n\in N,\forall m\in M,\forall t\in T \tag{16j}$$

Although the objective function explicitly focuses on minimizing latency and energy consumption, the quality of service aspects related to task completion are ensured through a set of constraints. Where constraint (16b) ensures that only one ES is selected by each UD for task offloading. (16c) indicates that the task follows a binary decision model, meaning that the task is either offloaded or computed locally. (16d) and (16e) guarantee that the UD transmission power and local computing capacity remain within their respective allocation budgets. (16f) ensures that the UD's battery energy stays between its minimum and maximum thresholds. (16g) specifies that the computation time for each task must not exceed the specified maximum time limit; otherwise, the task will be discarded. Tasks that exceed time or energy constraints incur significant penalties during optimization, thereby creating an implicit pressure to minimize such events. (16h) limits the number of UDs selecting a specific ES so that it does not exceed the ES's maximum user capacity. (16i) stipulates that a single subchannel can only be used by one task at any given time, and the number of tasks offloaded to the ES selected by the UD must not exceed the number of available subchannels. Finally, (16j) ensures that the total number of tasks offloaded to the selected ES does not exceed its storage capacity.

The optimization problem P1 is a typical non-convex Mixed-Integer Programming (MIP) problem, which is challenging to solve directly due to its NP-hard nature. According to the task model presented in Section 3.1, the offloading decision of the UD is made after selecting the appropriate ES. This implies an inherent coupling between offloading decisions and server selection. To simplify the problem, the original optimization problem P1 is decomposed into two subproblems: one focused on the selection decision between users and servers, and the other on the offloading decision. The optimization problem P2, which addresses the selection decision between users and servers, is formulated as follows:

$$\text{P2:}\quad \underset{\boldsymbol{y}}{minimize}\ \frac{1}{T}\sum_{t=1}^{T}\sum_{n=1}^{N}C_n(t) \tag{17a}$$

$$s.t.\quad \sum_{m\in M,n\in N} y_{n,m}(t)=1,\qquad \forall n\in N,\forall m\in M,\forall t\in T \tag{17b}$$

$$z_m(t)\le z_{max},\qquad \forall m\in M,\forall t\in T \tag{17c}$$

After obtaining selection decision $\boldsymbol{y}$ from optimization problem P2, we bring it into optimization problem P1, and we get offloading decision optimization problem P3 as shown below:

$$\text{P3:}\quad \underset{\boldsymbol{x},\boldsymbol{p},\boldsymbol{f}}{minimize}\ \frac{1}{T}\sum_{t=1}^{T}\sum_{n=1}^{N}C_n(t) \tag{18a}$$

$$s.t.\quad x_n(t)\in\{0,1\},\qquad \forall n\in N,\forall t\in T \tag{18b}$$

$$p^t_{min}\le p^t_{n,m}(t)\le p^t_{max},\quad \forall n\in N,\forall m\in M,\forall t\in T \tag{18c}$$

$$f^l_{min}\le f^l_n(t)\le f^l_{max},\qquad \forall n\in N,\forall t\in T \tag{18d}$$

$$b^h_{min}\le b^h_n(t)\le b^h_{max},\qquad \forall n\in N,\forall t\in T \tag{18e}$$

$$\varphi_n(t)\le\varphi_{max}(t),\qquad \forall n\in N,\forall t\in T \tag{18f}$$

$$\sum_{n\in Z_m} x_n(t)\le K,\qquad \forall n\in N,\forall m\in M,\forall t\in T \tag{18g}$$

$$\sum_{n\in Z_m} x_n(t)\beta_n(t)\le D_m,\quad \forall n\in N,\forall m\in M,\forall t\in T \tag{18h}$$

# 4 User-centric DRL Model Splitting Inference Scheme

To address the optimization problem P1 presented in Section 3, we propose a user-centric DRL model splitting inference scheme and a UCMS_MADDPG-based offloading algorithm. This transforms the problem of minimizing costs into a problem of maximizing returns. The proposed scheme proceeds as follows: First, a user-server co-selection algorithm is applied to address the matching and selection process between UDs and ESs. Following this, the UCMS_MADDPG algorithm is employed to jointly optimize offloading decisions, local computing power allocation, and transmission power budgeting for each UD, determining the optimal offloading strategy.

## 4.1 User-server Co-selection Algorithm

In the ES selection phase of traditional algorithms, the UD often prioritizes selecting the ES with maximum channel gain for task offloading. However, since both user capacity and storage capacity of the ES are considered in our system, traditional algorithms may lead to overload in some ESs, resulting in task loss and high offloading costs. Meanwhile, other ESs may be underutilized, wasting valuable computing resources. To obtain a better selection strategy, we propose a co-selection algorithm, where both the UD and the ES make decisions cooperatively to maximize their own interests and achieve high resource utilization. To implement this algorithm, we define a selection function to quantify the respective benefits of the UD and the ES.

From the UD's perspective, the primary objective is to minimize its own computation delay and energy consumption. When selecting the target ES, the UD prefers those offering faster transmission rates, shorter execution delays, and more available computing resources. Consequently, the total offloading delay and offloading energy consumption can be directly determined. Then the selection function of UD$n$ is expressed as $I_m(t)$ and the following equality is satisfied:

$$I_m(t) = \sum_{n \in Z_m} \left( \rho_1 \tau_{n,m}^a(t) + \rho_2 e_{n,m}^o(t) \right), m \in M \tag{19}$$

From the ES's perspective, its objective is also to minimize computation delay. Therefore, the ES will prioritize UDs with smaller computation tasks to complete the processing more quickly. The selection function of ES$m$ is denoted as $I_n(t)$ and the corresponding equality is defined as follows:

$$I_n(t) = \tau_{n,m}^{a}(t), n \in N \tag{20}$$

The user-server co-selection algorithm is mainly determined according to the selection function and the user capacity of the ES, and the specific implementation is shown in Algorithm 1.

---

**Algorithm 1:** User-server Co-selection Algorithm

---

**Input:** $I_m(t)$, $I_n(t)$, $z_{max}$

**Output:** $y_{n,m}(t)$, $Z_m$

Initialize the UD set *N*, the ES set *M*, the rejection list $N^r(t) = N$ , and initialize the application list $N_{n,m}^{a}(t)$

**for** each ES*m* **do**

  Initialize the selection list $Z_m(t)$ , user capacity $z_m(t)$

**end for**

**while** $N^r(t) > 0$ **do**

  **for** each UD$n \in N^r(t)$ **do**

    **if** $0 < z_m(t) \le z_{max}$ of the ES*m* **then**

      Sort the selection function values of all ES*m* in ascending order

      Send the request to the ES*m* with the lowest selection function value

      Update application list $N_{n,m}^{a}(t)$

    **else**

      **break**

    **end if**

  **end for**

  **for** each ES*m* **do**

    **if** $z_m(t) = z_{max}$ of the ES*m* **then**

      Skip the ES*m*

      Sort the selection function values of all UD*n* in $N_{n,m}^{a}(t)$ in ascending order

      **for** each UD*n* after sorting **do**

        **if** $z_m(t) < z_{max}$ of the ES*m* **then**

          The UD*n* is selected by the ES*m*

          Updating selection list $Z_m(t) = Z_m(t) + \mathrm{UD}n$

          Updating rejection list $N^r(t) = N^r(t) - \mathrm{UD}n$

        **else**

          **break**

        **end if**

      **end for**

    **end if**

    Clear application list $N_{n,m}^{a}(t)$

```
            Update the user capacity of the ESm z_m(t) = z_m(t) + UDn
        end for
    end while
```

First, the rejection list $N^r(t)$ is initialized, containing all UDs. An application list $N^a_{n,m}(t)$ is used to track each UD's application to the ES, while the selection list $Z_m(t)$ for each ES is initialized to store the connected UDs and track the current user capacity $z_m(t)$. The algorithm then enters an iterative process. In each iteration, the unassigned UDs are sorted in ascending order according to the ES selection function $I_n(t)$, and each sends an application to the ES with the lowest function value. Upon receiving applications, each ES sorts the requesting UDs in ascending order according to the UD selection function $I_m(t)$, and accepts UDs with the lowest function values until its user capacity limit is reached. The selected UDs are added to the selection list, and the remaining requests are rejected. After each selection round, the ES clears its application list and updates its available user capacity. The iteration ends when the rejection list becomes empty; otherwise, the process continues until all UDs in the rejection list are successfully assigned to appropriate ESs.

It is important to note that the joint optimization problem P1 is NP-hard. The decoupling into subproblems P2 and P3 is a pragmatic approximation to achieve a tractable solution. Algorithm 1 addresses P2 heuristically by establishing a beneficial initial matching between UDs and ESs. The selection functions $I_m(t)$ and $I_n(t)$ are designed to substitute for the key components of the system cost. Algorithm 1 effectively reduces the complexity of the subsequent offloading decision problem P3 and provides a high-quality initialization for the UCMS_MADDPG algorithm. The overall effectiveness of this decomposed strategy is empirically validated by the performance results presented in Section 5.

## 4.2 MDP Formulation

DRL is based on the Markov decision process. Before designing a DRL algorithm, the research problem must be modeled and converted into an MDP framework. To solve the optimization problem P3 in Section 3.5, this section models the task offloading decision $\boldsymbol{x}$, local computing power allocation $\boldsymbol{f}$, and transmission power budget allocation $\boldsymbol{p}$ of the UD as an MDP. The three key components of the MDP are described in detail below.

### 4.2.1 State

At each time slot $t$, the system environment provides the current state, which consists of the task parameters of UD$n$ $\left(\beta_n(t), c_n(t), \varphi_n(t)\right)$, local computing power $f_n^l(t)$, offloading transmission power $p_{n,m}^t(t)$, battery energy $b_n(t)$, and normalized channel gain $g_{n,m}(t)$ between UD$n$ and each edge server ES$m$. This state is expressed as $S_n(t)$:

$$S_n(t) = \left\{s_n^t(t), s_n^f(t), s_n^p(t), s_n^b(t), s_n^g(t)\right\} \in S(t) \tag{21}$$

where task state is defined as $s_n^t(t) = \left(\beta_n(t), c_n(t), \varphi_n(t)\right)$, local computing power state as $s_n^f(t) = f_n^l(t)$, offloading transmission power state as $s_n^p(t) = p_{n,m}^t(t)$, battery energy state as $s_n^b(t) = b_n(t)$, and normalized channel gain state as $s_n^g(t) = g_{n,m}(t)$.

### 4.2.2 Action

At time slot $t$, the system's computing tasks involve three decision variables: binary offloading decision $x_n(t)$, local computing power allocation $f_n^l(t)$, and transmission power allocation $p_{n,m}^t(t)$. In the user-centric model splitting inference scheme, the UD$n$ first calculates the corresponding offloading pre-decision and resource allocation for its CPU (first-stage model inference), then transmits its status and action information to the selected edge server ES$m$. The ES$m$ subsequently uses its CPU to calculate the offloading pre-decision (second-stage model inference). The execution process is as follows:

(1) For a UD$n$ that makes a local computing decision, the ES$m$ respects this decision.

(2) For a UD$n$ that makes an offloading computing decision, the ES$m$ makes a hybrid decision based on the recommendation of the UD$n$ and its resources.

(3) If the number of UDn requesting offloading exceeds the available subchannels

of ESm, or if the total size of the offloaded tasks surpasses the storage capacity of ESm, the ES*m* selectively approves offloading requests based on the hybrid decision strategy.

(4) If both the number of offloading UDs and the total task size remain within the resource constraints of ES*m*, the ES*m* approves all offloading requests. The corresponding tasks are then assigned to subchannels, and offloading computation begins.

The entire action space is denoted as $A(t)$ and is split into two distinct components: the UD action and the ES action. In each time slot *t*, every UD*n* generates three continuous actions in the range [0,1] which represent the offloading pre-decision and resource allocation. These actions form the UD action set, represented as $A_u(t)$:

$$A_u(t) = \left\{ x_n^u(t), f_n^u(t), p_{n,m}^u(t) \right\} \tag{22}$$

where $x_n^u(t)$ represents the pre-decision offloading action of the UD*n*. If $x_n^u(t) < 0.5$, let $x_n(t) = 0$ in equation (1). Otherwise, it is transmitted to the ES*m* for hybrid decision. The local computing power allocation $f_n^l(t)$ is calculated according to the $f_n^u(t)$ action, as shown in equation (23). Similarly, the transmission power allocation $p_{n,m}^t(t)$ is calculated according to the action $p_{n,m}^u(t)$, as shown in equation (24).

$$f_n^l(t) = \max\left( f_{min}^l, f_n^u(t) f_{max}^l \right) \tag{23}$$

$$p_n^t(t) = \max\left( p_{min}^t, p_{n,m}^u(t) p_{max}^t \right) \tag{24}$$

For any UD*n* with $x_n^u(t) \geq 0.5$, the UD*n* submits an offloading request to the selected ES*m*. After obtaining its status and action information, the ES*m* makes a binary hybrid decision to determine which requests are approved. The ES action is denoted as $A_m(t)$, as follows:

$$A_m(t) = \left\{ x_n^m(t) \right\} \tag{25}$$

where let $x_n(t) = 1$ for the UD*n* of consent application and $x_n(t) = 0$ otherwise. The binary decision $x_n^m(t)$ is determined by the ES*m* based on an evaluation of a fixed-dimensional input vector. This vector incorporates the UD's state $S_n(t)$, its pre-decision action $A_u(t)$, and a summary of the current global resource status of the ES.

#### 4.2.3 Reward

After the UD$n$ executes its action, the system environment is updated, and reward feedback is obtained. The reward function plays a critical role in DRL, making it essential to fully incorporate the constraints from the optimization problem. While other constraints for the UD and ES have been addressed in previous works, here we focus on the task time and battery energy constraints. We use the inverse of the cost function from Section 3.5 as the reward term, and use the task time and battery energy limits of UD$n$ as the penalty term.

In[15], the reward for battery depletion included a drop penalty. In this paper, we define that the penalty starts when the available energy falls below the specified minimum power threshold. The penalty function is denoted as $P_n(t)$ and satisfies the following:

$$P_n(t) = \rho_1 \min\left(\left(\varphi_n(t) - \tau_n(t)\right), 0\right) + \rho_2 \min\left(\left(b_n(t) - b_{min}^h\right), 0\right) \tag{26}$$

Therefore, the reward function of each UD$n$ within the shared time slot $t$ is defined as $r_n(t)$:

$$r_n(t) = -\left(\frac{1}{N}\sum_{n=1}^{N} C_n(t) + P_n(t)\right) \tag{27}$$

### 4. 3 UCMS_MADDPG-based Offloading Algorithm

According to the MDP transformation model in Section 4.2, this chapter proposes an offloading algorithm using UCMS_MADDPG. The UCMS_MADDPG algorithm implicitly realizes a model-splitting architecture through its hierarchical decision structure. In this paradigm, the user pre-decision corresponds to the initial inference stage of a split model, where local computations are performed based on device-specific states. The server hybrid decision subsequently completes the final inference stage, integrating global system information to refine decision outcomes.

The algorithm is designed on the actor-critic framework of the MADDPG algorithm, combined with DQN. In the AIoT multi-edge server deployment environment, each UD$n$ is treated as an agent. The policy network (Actor) of each agent takes the local state as input, while the Q-network (Critic) takes the global state and the set of actions from all agents as input. Additionally, we introduce a priority sampling mechanism for reward error trade-off, which more effectively determines the priority of samples during the experience replay process.

Fig. 3 shows the architecture diagram of the UCMS_MADDPG-based offloading algorithm. After the UD$n$ generates the output, the following operations will be performed: if $x_n^u(t) < 0.5$, let $x_n(t) = 0$, and start the local computation. Otherwise, it transmits $x_n^u(t), f_n^u(t), p_{n,m}^u(t)$ to the ES$m$ for hybrid decision. Subsequently, the ES$m$ produces binary decisions and feeds them back to the UD$n$. Finally, the reward within the shared time slot $t$ is calculated and provided to the ES$m$ for training. The UD$n$ is trained using a trade-off value of reward and TD error computed by the ES$m$.

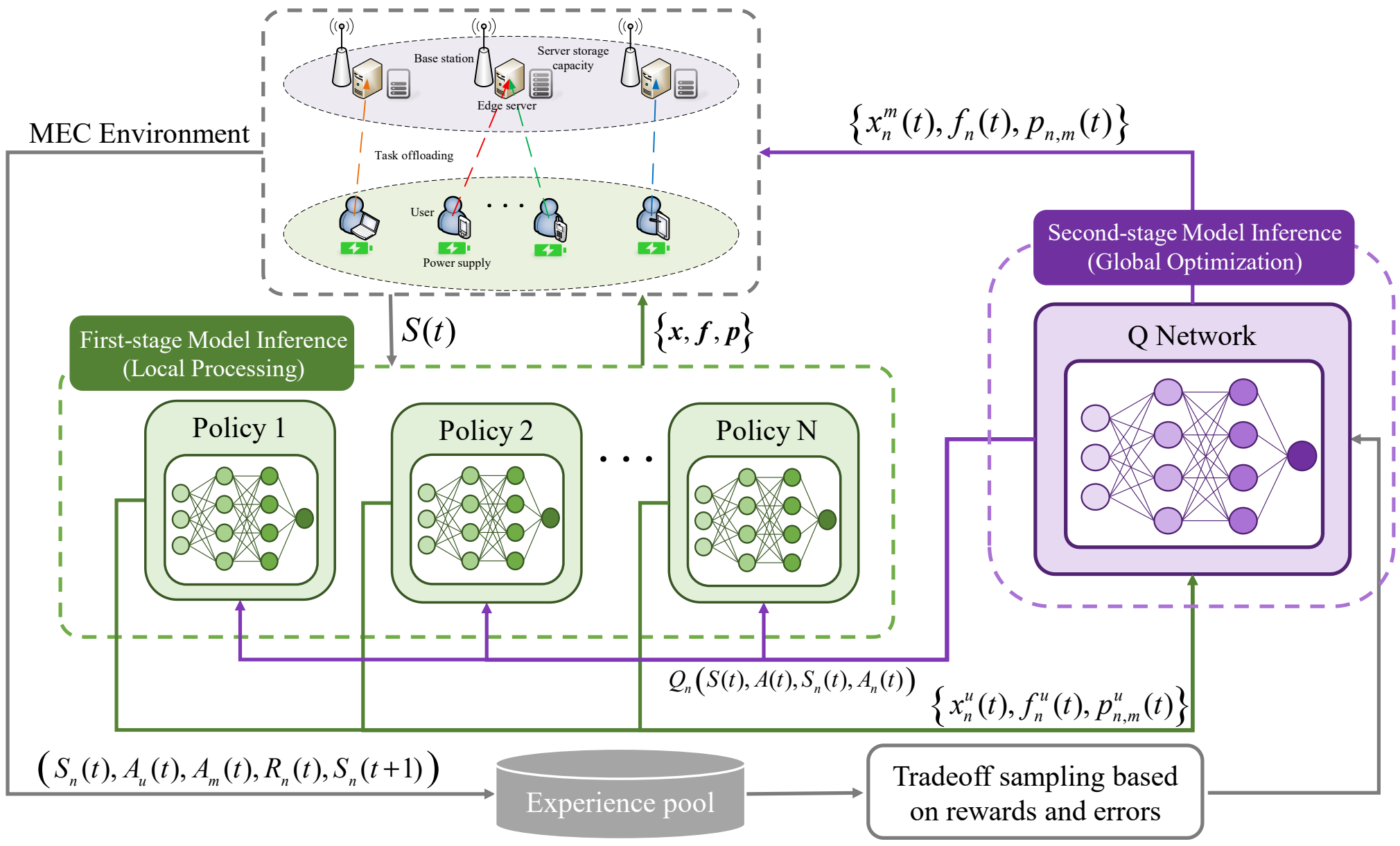

**Fig.3 Offloading algorithm architecture based on UCMS_MADDPG**

A fundamental aspect of the UCMS_MADDPG algorithm design is the clear separation between neural network-based value estimation and rule-based, resource-aware decision-making. The algorithm relies on a centralized critic network with a fixed input dimension, which evaluates global state-action pairs across all UDs in the system. Algorithm 1 influences the content of the UD states, while the dimensions of these states remain fixed. The ES does not possess independent policy networks. Instead, their binary offloading decisions are generated by a decision module that interprets the Q-values provided by the fixed-dimension critic in the context of the ES's instantaneous resource constraints. This architecture ensures that the neural network training process remains stable and is not subject to dynamic changes in the number of UD associated with each ES.

### 4.3.1 Preferential Sampling Mechanism of Reward Error Trade-off

In the experience replay process, the existing preferential sampling mechanism typically uses the TD error as an index to measure the importance of experience samples. However, due to the continuous update of neural network parameters, the TD error changes dynamically. As a result, some samples with gradually decreasing TD errors may still be selected with high probability for training the Q-network. This can lead to overfitting, causing the agent to focus too much on these samples and potentially fall into a local optimum, which negatively impacts the global optimization of the policy. To address this issue, we propose using a trade-off between the current reward and the TD error to update the sampling priority of experience samples. This composite priority mechanism effectively balances short-term feedback and long-term error, improving the diversity of experience replay and enhancing training robustness. Ultimately, it helps the agent escape local optima and achieve better policy performance.

So we use the trade-off value of current reward $r_n(t)$ and TD error $\delta_n(t)$ to calculate the composite priority of experience samples, denoted as $\eta(t)$. Firstly, $\eta_r(t)$ is defined as the priority based on the current reward and $\eta_\delta(t)$ is defined as the priority based on the TD error, which satisfies the following two equalities.

$$\eta_r(t) = |r_n(t)| + \varepsilon \tag{28}$$

$$\eta_\delta(t) = |\delta_n(t)| + \varepsilon \tag{29}$$

Define $\upsilon_r$ and $\upsilon_\delta$ as the trade-off factors of $\eta_r(t)$ and $\eta_\delta(t)$ respectively. They satisfy $\upsilon_r + \upsilon_\delta = 1$. The compound priority $\eta(t)$ is calculated as follows:

$$\eta(t) = \upsilon_r \eta_r(t) + \upsilon_\delta \eta_\delta(t) \tag{30}$$

Assuming that the set of experience samples selected in each batch is $\mathrm{B}$ and the number of experience samples selected is $N_\mathrm{B}$, the sampling probability of experience sample $i$ ($i \le N_\mathrm{B}$) is represented by $p_i^\mathrm{B}(t)$:

$$p_i^\mathrm{B}(t) = \frac{\eta_i(t)}{\sum_{i \in \mathrm{B}} \eta_i(t)} \tag{31}$$

### 4.3.2 Policy Updates

Since the objective of this paper is to design an offloading algorithm that maximizes the UD's returns over a continuous period, it must consider both immediate rewards and long-term cumulative returns using Bellman's equation. Given the immediate reward $R_n(t) = r_n\left(S(t), A(t)\right)$, the Q function satisfies the following:

$$\begin{aligned} &Q_n\left(S(t), A(t), S_n(t), A_n(t)\right) = \\ &\mathbb{E}\left[R_n(t) + \gamma P_n\left(S(t+1)\right) \max_{S_n(t+1), A_n(t+1)} Q_n\left(S(t+1), A(t+1), S_n(t+1), A_n(t+1)\right)\right] \end{aligned} \quad (32)$$

where $\gamma$ is the discount factor, $S(t)$, $A(t)$ and $S(t+1)$, $A(t+1)$ are the set of the current state and action, and the next state and action, respectively, of all UD$n$ that select the ES$m$. $S_n(t)$, $A_n(t)$ and $S_n(t+1)$, $A_n(t+1)$ are the sets of the current state and action, and the next state and action, respectively, of all UD$n$ in the hybrid decision of the ES$m$.

In the training framework, a centralized Q network is employed to estimate the Q-value. The input to this Q network is the fixed-dimensional concatenation of states and actions from all UDs in the system, which ensures a consistent input size during training. The policy networks are associated exclusively with the UDs. We define that UD action $A_u(t)$ mentioned in Section 4.2.2 follows directly from the current state $S_n(t)$, i.e., $A_u(t) \leftarrow \lambda\left(S_n(t)\right)$, where $\lambda$ is a policy parameter of the UD. ES action $A_m(t)$ is derived based on the UD action, i.e., $A_m(t) \leftarrow \theta\left(S(t), A(t), S_u(t), A_u(t)\right)$, where $S(t)$ is the set of states of all UD$n$ that select this ES$m$, $A(t)$ is the set of actions, and similarly, $S_u(t)$ and $A_u(t)$ are the set of states and actions of all UD$n$ with $x_n^u(t) \geq 0.5$. $\theta$ is policy parameter of all ESs. The ES's decision parameter, which produces the binary actions $A_m(t)$, operates on the outputs of the centralized Q network and follows deterministic rules that account for resource constraints.

The policy update procedure consists of two components: Q network updating and policy network updating. During training, a mini-batch of data $\mathrm{B}$ ($i \in \mathrm{B}$) is randomly sampled from the experience pool of the preferential sampling mechanism of reward error trade-off. First, training on ES learns the offloading strategy by maximizing the global reward and uses the target Q network $Q'_{i,n}$ to calculate the target Q value, which is represented by $y_i(t)$ :

$$y_i(t) = R_i(t) + \gamma \max_{S_{i,n}(t+1), A_{i,n}(t+1)} Q'_{i,n}(t+1) \tag{33}$$

Then, the loss function of the Q network $Q_{i,n}$ is defined according to the Weighted Mean Squared Error (Weighted MSE), and the loss function is defined as $L\left(Q_{i,n}\right)$:

$$L\left(Q_{i,n}\right) = \mathbb{E}\left[\eta(t)\left(y_i(t) - Q_{i,n}(t)\right)^2\right] \tag{34}$$

And update the Q network parameter $\theta_{Q_{i,n}}$ as $\theta_{Q_{i,n}} \leftarrow \theta_{Q_{i,n}} - \alpha_Q \nabla_{\theta_{Q_{i,n}}} L\left(Q_{i,n}\right)$, where $\alpha_Q$ is the Q network learning rate. To improve the stability of network training, the soft update mechanism is used to synchronize the parameters $\theta_{Q'_{i,n}} \leftarrow \omega\theta_{Q_{i,n}} + \left(1-\omega\right)\theta_{Q'_{i,n}}$ of the target Q network $Q'_{i,n}$, where $\omega$ is the target network update coefficient.

The policy network is updated through gradient ascent to maximize expected reward. For each training sample, the policy network relies on the target Q value provided by the Q network to perform its own network update and policy optimization, thus ensuring that the global policy optimization proceeds in a consistent direction. The gradient of the policy network $\mu_n$ is denoted as $\nabla_{\lambda_{\mu_n}} J(\mu_n)$:

$$\nabla_{\lambda_{\mu_n}} J(\mu_n) = \mathbb{E}\left[\nabla_{\lambda_{\mu_n}} \mu_n\left(S_i(t)\right) \nabla_{A_{i,n}(t)} Q_i(t)\right] \tag{35}$$

The policy network parameter is then updated as $\lambda_{\mu_n} \leftarrow \lambda_{\mu_n} + \alpha_\mu \nabla_{\lambda_{\mu_n}} J\left(\mu_n\right)$, where $\alpha_\mu$ is the policy network learning rate. Finally, the target policy network $\mu'_n$ is soft updated as $\lambda_{\mu'_n} \leftarrow \omega\lambda_{\mu_n} + \left(1-\omega\right)\lambda_{\mu'_n}$ .

In summary, the specific steps of UCMS_MADDPG-based offloading algorithm are shown in Algorithm 2.

---

**Algorithm 2:** UCMS_MADDPG-based Offloading Algorithm

---

**Input:** Maximum number of rounds $Ep_{max}$ , Number of training rounds $Ep_{train}$

**Output:** $x_n(t)$, $f_n^l(t)$, $p_{n,m}^t(t)$

1: **while** $Ep_{now} < Ep_{max}$ **do**

2: Execute the reset() method to reset the system environment

3: Get the initial state of the current environment $S_n(t)$

4: **for** $t = 1 \ldots T$ **do**

**for** each UD$n$ **do**
Obtain offloading action based on local status $A_u(t), A_m(t)$
**end for**
Get the reward $R_n(t)$, and the next state $S_n(t+1)$
Save $\left(S_n(t), A_u(t), A_m(t), R_n(t), S_n(t+1)\right)$ to the experience replay buffer
Execute the step() method to update the state $S_n(t) \leftarrow S_n(t+1)$
**end for**
**for** $Ep_{train}$ **do**
The set of experience samples collected from the experience replay buffer is B
Extract the data $\left(S_n(t), A_u(t), A_m(t), R_n(t), S_n(t+1)\right)$ of each batch sample B
**for** each UD$n$ **do**
The goal policy network is used to calculate the next goal action $A_u(t+1)$
**end for**
**for** B **do**
**if** $x_n^m(t) = 1$ **then**
Execute Equation (33) to calculate the target Q value $y_i(t)$
**else if** $x_n^m(t) = 0$ **then**
The default Q value is computed using zero states and actions
**end if**
Update the current Q value of the Q network
Cumulative reward error trade-off values for prioritized experience replay
**end for**
Execute Equation (34) to calculate the Q network loss $L\left(Q_{i,n}\right)$
Update the Q network parameters $\theta_{Q_{i,n}}$
Soft update target Q network parameters $\theta_{Q'_{i,n}}$
**for** each UD$n$ **do**
Generate new actions using a policy network $A_u(t)$
Calculate Q value related to new actions using the Q network of the ES
Execute Equation (35) to calculate policy network gradient $\nabla_{\lambda_{\mu_n}} J(\mu_n)$, and update $\lambda_{\mu_n}$
Soft update target policy network $\lambda_{\mu'_n}$
**end for**
Update the sampling weights of samples based on the composite priority $\eta(t)$
**end for**
**end while**

# 5 Simulation Results and Analysis

## 5.1 Simulation Parameter Setting

In the AIoT multi-edge server deployment dynamic MEC environment established above, the number of UD $N$ is set to 48, and the number of MEC servers $M$ is set to 3. The parameter settings from[15, 21] were used to construct the experimental environment. All UDs have identical local computing capacity, transmission power allocation, and battery threshold. While modern ESs typically have storage capacities of several GB or more, we chose a smaller value to simulate high-load conditions, setting the server storage capacity to 400 MB. Specific parameter settings are provided in Table 3.

The proposed UCMS_MADDPG algorithm is implemented using the PyTorch framework. Simulations were conducted with Python 3.9 and PyTorch 2.2 on a host machine equipped with an NVIDIA GeForce RTX 4070 GPU and a 13th Gen Intel Core i5-13600KF CPU (running at 3.5GHz) and 32GB of RAM. The strategy network and Q network are both composed of two fully connected layers, with 64 and 512 neurons, respectively. The learning rates of the policy network and Q network are set to $1\times e^{-4}$ and $1\times e^{-3}$ respectively. The number of experience samples in each batch of training $N_{\mathrm{B}}$ is 64, the capacity of the experience buffer pool is $1\times e^{5}$, the discount coefficient is 0.99, and the Adam optimizer is used for gradient descent calculation during training.

**Table 3 System Parameter Setting**

| Parameter | Value |
|---|---|
| The number of edge servers $M$ | 3 |
| The number of user devices $N$ | 48 |
| Maximum tolerated delay $\varphi_{max}(t)$ | [0.1,1] s |
| Task size in bytes $\beta_n(t)$ | [1-50] MB |
| The number of CPU cycles required by the task $c_n(t)$ | [300,700] cycles |
| Maximum local computing capacity $f^{l}_{max}$ | 1.5 GHz |
| Minimum local computing power $f^{l}_{min}$ | 0.4 GHz |
| Maximum transmission power $p^{t}_{max}$ | 24 dBm |
| Minimum transmission power $p^{t}_{min}$ | 1 dBm |
| Maximum battery threshold $b^{h}_{max}$ | 3.2 MJ |
| Minimum battery threshold $b^{h}_{min}$ | 0.5 MJ |
| Maximum user capacity $z_{max}$ | 16 |
| The number of CPUs equipped $U_m$ | 8 |
| Maximum storage capacity $D_m$ | 400 MB |
| Server CPU computing power $f_m^s$ | 4 GHz |
| Normalized channel gain $g_{n,m}(t)$ | [5,14] dB |
| Energy consumption factor $\kappa$ | $5\times e^{-27}$ |
| Wireless Network bandwidth $W$ | 40 MHz |
| The number of subchannels $K$ | 10 |
| Harvested energy $d_n(t)$ | $1\times e^{-3}$ J |
| Cost weight coefficient $\rho_1$ | 0.5 |
| Cost weight coefficient $\rho_2$ | 0.5 |

## 5. 2 Baseline Comparison Scheme

To assess the performance of the UCMS_MADDPG algorithm, we compare it against several benchmark algorithms based on MADDPG through simulation experiments:

- RD_UCMS_MADDPG: Only the UCMS_MADDPG algorithm is used to optimize resource and offloading decisions. User or server selection is performed randomly, without any co-selection mechanism.
- MADDPG: Extends the DDPG algorithm by using centralized policy evaluation and decentralized policy execution, allowing multiple agents to learn collaboratively. When updating the policy, the Q network utilizes global information to guide the policy network training. When interacting with the environment, the policy network generates actions only by acquiring local states.
- OFFLOADCOSTFIRST_MADDPG: A heuristic MADDPG algorithm with minimum offloading cost first. By calculating and ranking the combined cost of each task, the solution with the lowest offloading cost is given the highest priority.
- DEADLINEFIRST_MADDPG: A heuristic MADDPG algorithm with a deadline first. By calculating and sorting the deadlines of all tasks, the task with the earliest deadline is given the highest priority.

## 5. 3 Algorithms Performance Analysis

First, we set up 2000 training episodes, each containing 10 time slots, and conducted the experiment 10 times to generate the results. Fig. 4 shows the convergence comparison between UCMS_MADDPG and RD_UCMS_MADDPG in the training environment. Initially, the reward values of both algorithms are relatively low. However, as training progresses, the reward values increase and eventually stabilize at higher levels. It can be seen that UCMS_MADDPG converges faster and gradually stabilizes after around 60 training rounds, indicating that the UCMS_MADDPG network has been effectively trained. In contrast, RD_UCMS_MADDPG converges more slowly, with a lower overall system return compared to UCMS_MADDPG. The superior convergence performance of UCMS_MADDPG can be attributed to two key factors. The user-server co-selection algorithm provides a stable foundation for learning by establishing efficient initial associations. The priority sampling mechanism based on reward-error trade-off ensures more effective experience replay, accelerating the learning process while maintaining stability. This combination enables our method to quickly identify promising policies.

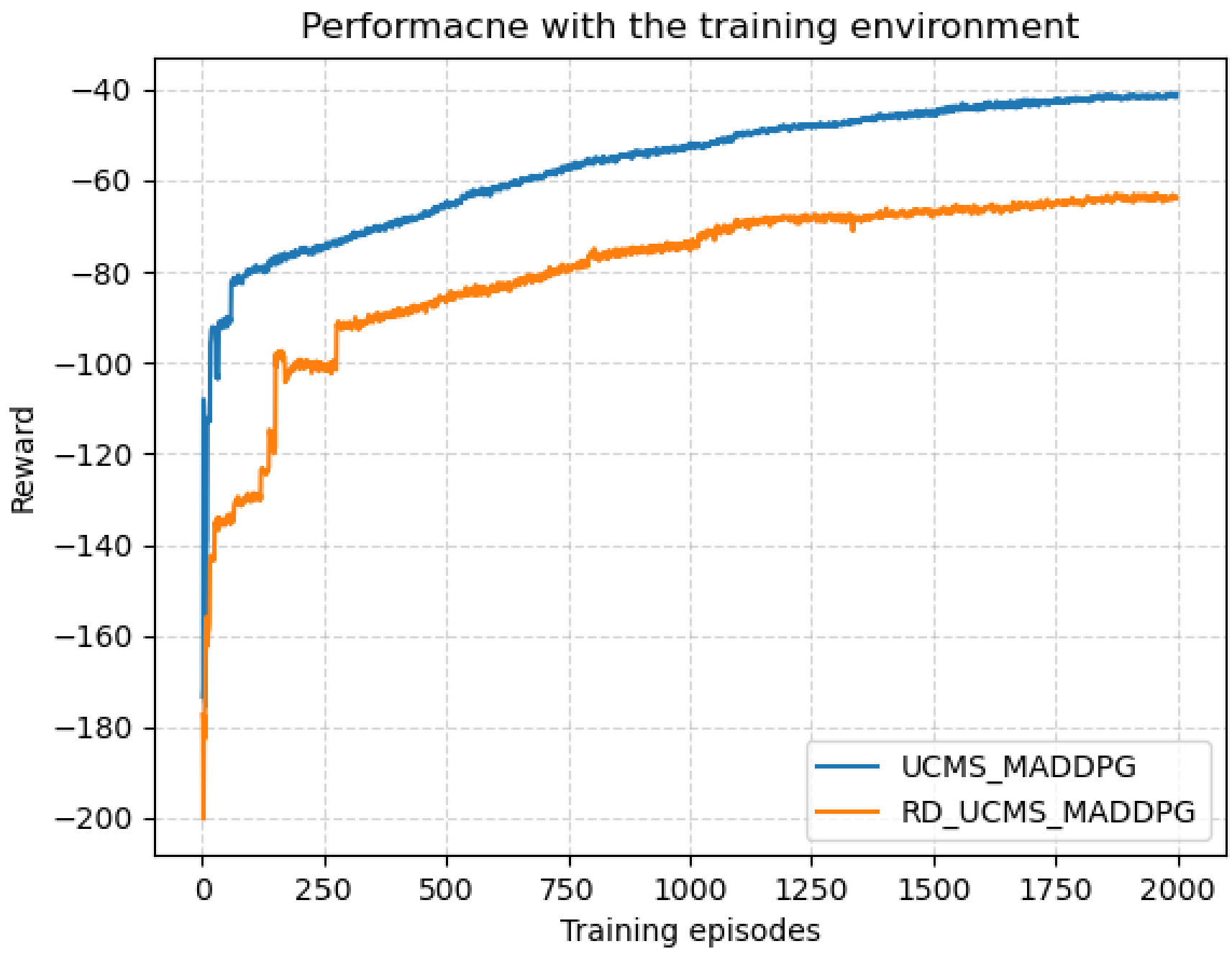

**Fig.4 Comparison of convergence performance in the training environment**

Fig. 5 shows the performance comparison between UCMS_MADDPG and four baseline methods: RD_UCMS_MADDPG, MADDPG, OFFLOADCOSTFIRST_MADDPG, and DEADLINEFIRST_MADDPG. It is evident that UCMS_MADDPG outperforms all other methods, effectively utilizing global information and avoiding local optima. RD_UCMS_MADDPG exhibits larger reward fluctuations and lower reward values, suggesting that random user-server selection is not suitable for dynamic offloading scenarios. MADDPG consistently produces lower average rewards, as it relies solely on the actor to analyze information, with the critic only providing feedback. This structure leads to longer convergence times in complex environments. In contrast, OFFLOADCOSTFIRST_MADDPG and DEADLINEFIRST_MADDPG introduce artificial priority rules, which improve early-stage efficiency. However, these rules reduce the algorithm's flexibility, causing the heuristic methods to perform worse than UCMS_MADDPG in dynamic task and resource environments, and leading to local optima early in training. While they achieve better returns than MADDPG, they require more rounds to converge.

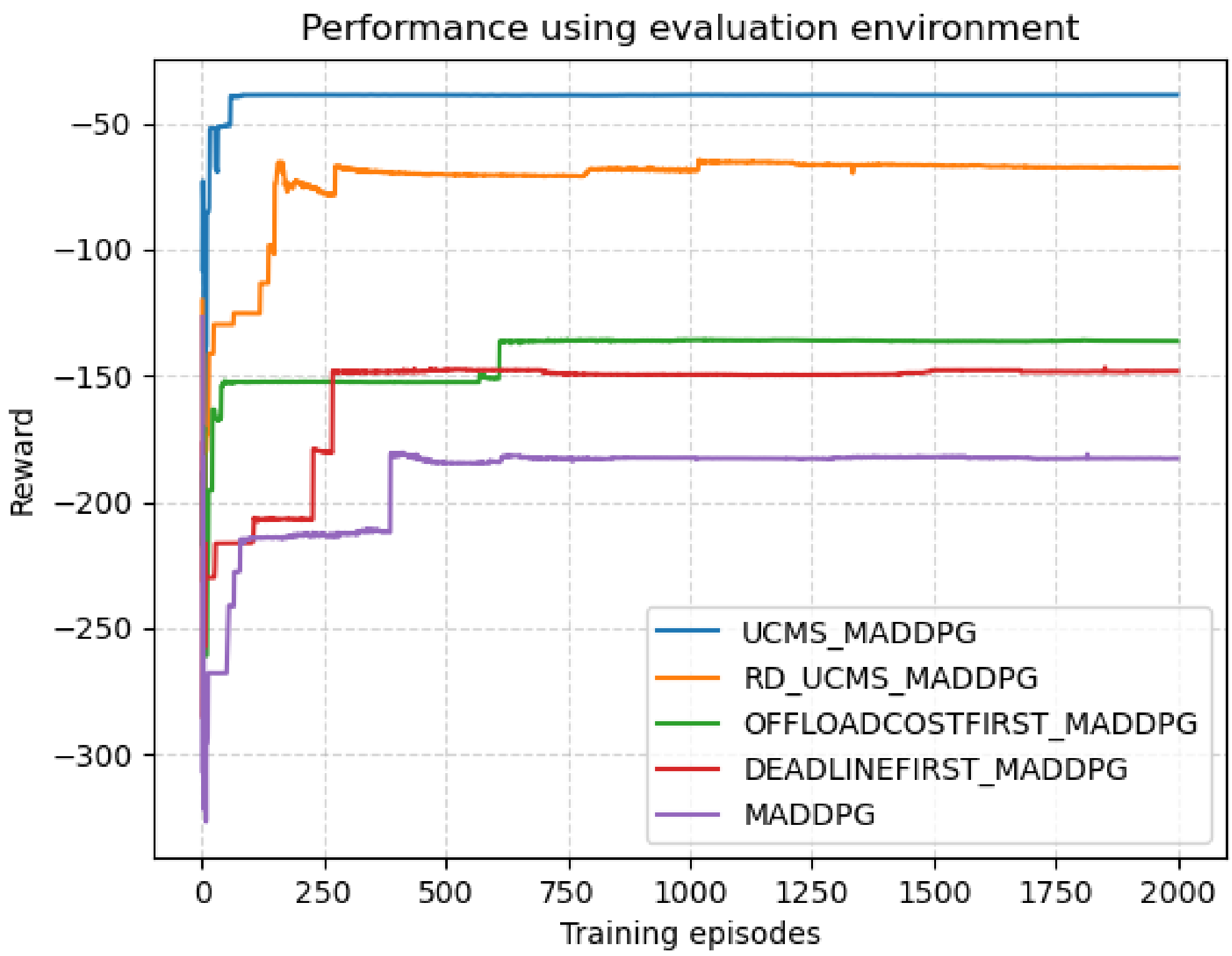

**Fig.5 Performance comparison for different algorithms in the evaluation environment**

To further evaluate the performance of the UCMS_MADDPG-based offloading algorithm, we compare it with the four benchmark algorithms in terms of total system cost as the number of UDs increases from 12 to 57, as illustrated in Fig. 6. As the number of UDs grows, the ES load increases, leading to a decrease in available computation and communication resources per UD, which results in a significant rise in total cost. Overall, MADDPG incurs the highest total cost. This is followed by OFFLOADCOSTFIRST_MADDPG and DEADLINEFIRST_MADDPG, while UCMS_MADDPG consistently maintains the lowest total system cost. This demonstrates the superior adaptability of UCMS_MADDPG in high-load offloading scenarios.

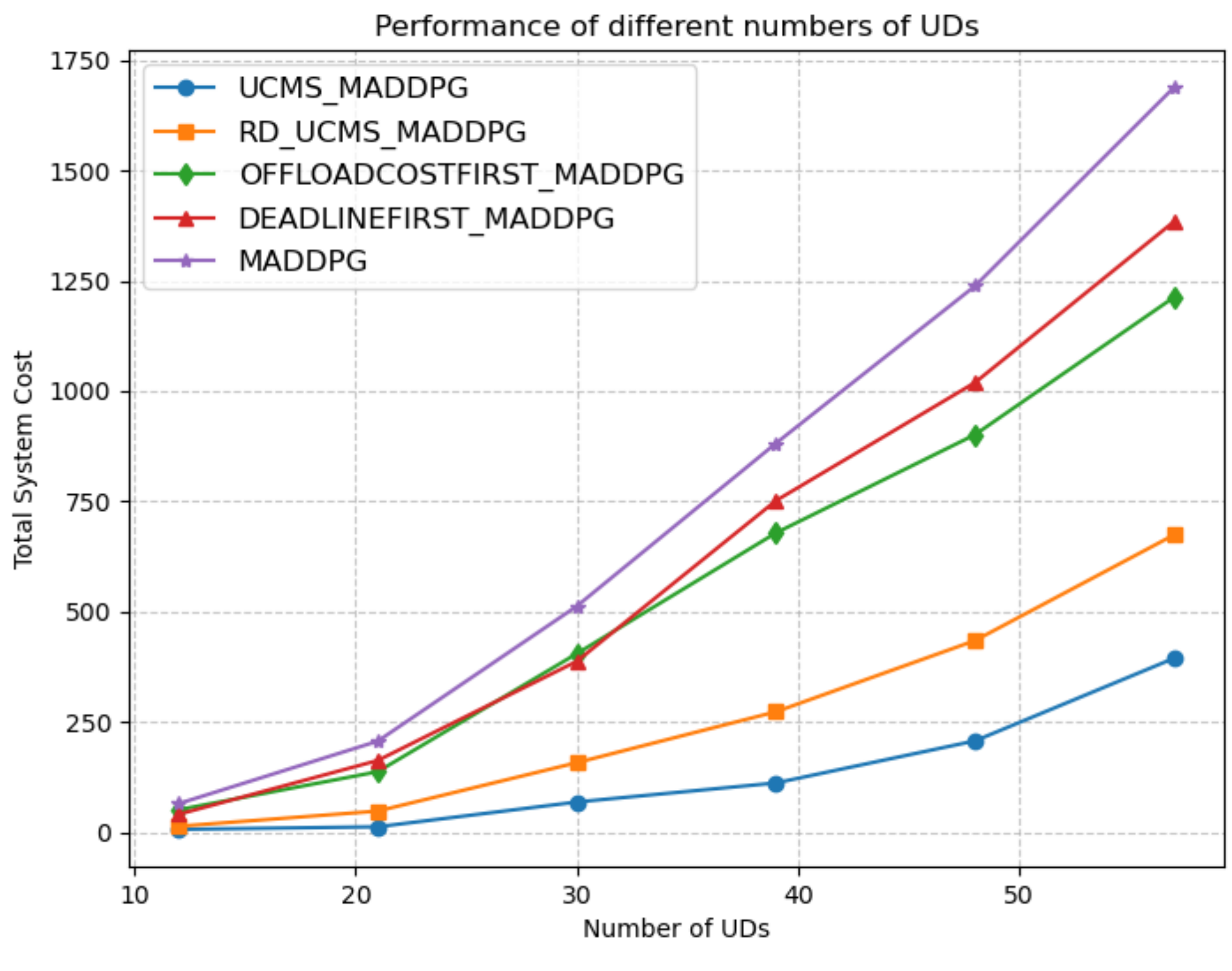

**Fig.6 Comparison of total system cost with different numbers of UDs**

We then further compare the performance of UCMS_MADDPG with the four benchmark algorithms in terms of server participation decisions. To reduce the complexity of image information, we apply the Savitzky-Golay filter to smooth the data and down-sample every 100 rounds, as shown in Fig. 7. The figure reveals that UCMS_MADDPG maintains a higher server participation rate, which correlates with its lower total system cost. In contrast, MADDPG and the two heuristic algorithms show a high server participation rate initially, but this rate gradually decreases with each iteration and eventually stabilizes at a low level. This suggests that MADDPG, when used alone, struggles to effectively adapt to the complexities of resource competition and task allocation in dynamic scenarios. The consistent outperformance of UCMS_MADDPG across different metrics stems from its comprehensive approach to addressing the joint optimization problem. While heuristic methods like OFFLOADCOSTFIRST_MADDPG and DEADLINEFIRST_MADDPG prioritize single factors, our method dynamically balances multiple objectives through policy learning. The integration of model splitting inference enables more sophisticated decision-making, adapting to both user requirements and server constraints, which results in more balanced resource utilization and improved overall performance.

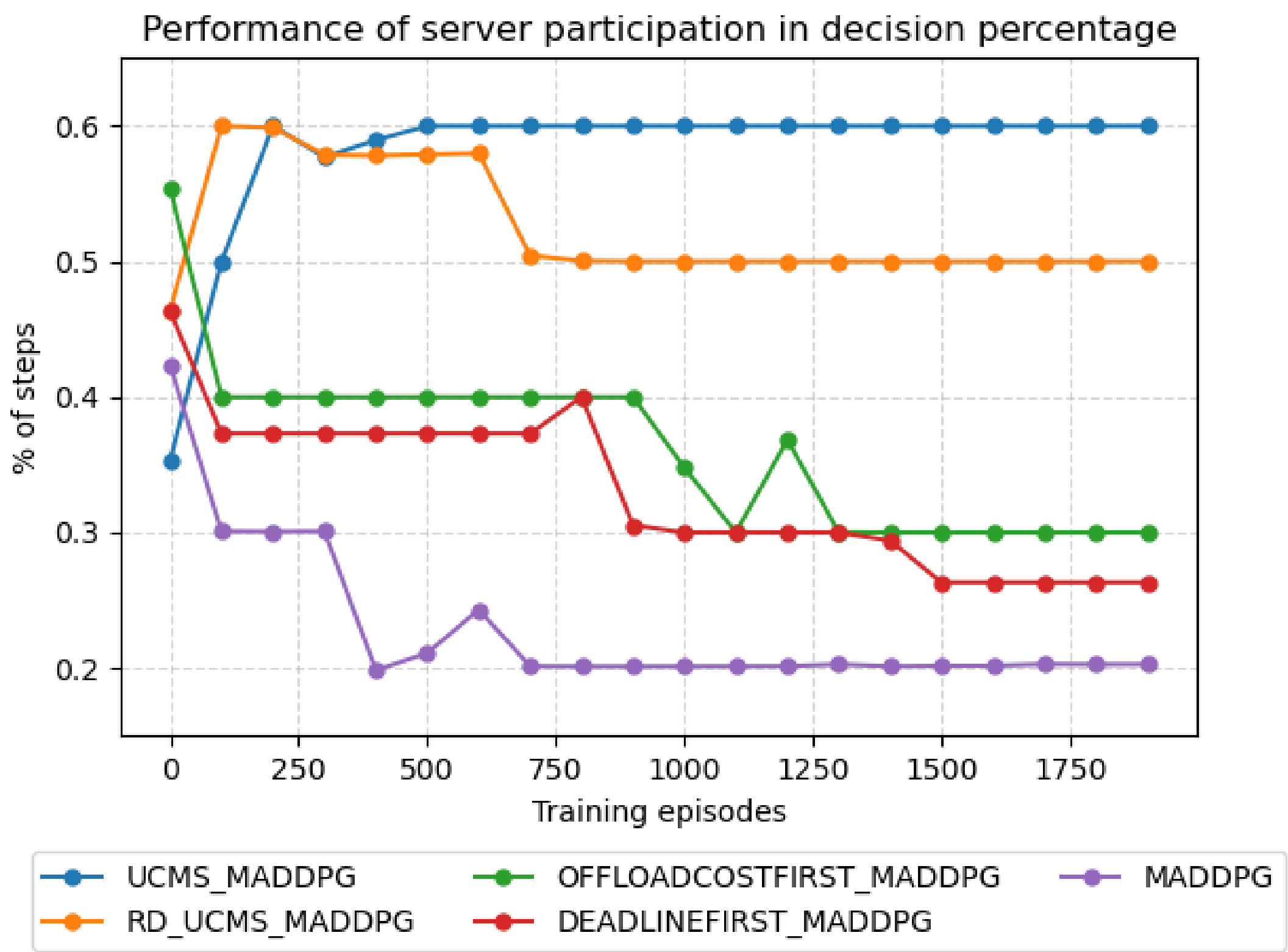

**Fig.7 Percentage of servers participating in decision for different algorithms**

Similarly, Fig. 8 compares the performance of UCMS_MADDPG with the four benchmark algorithms in terms of task timeouts. UCMS_MADDPG demonstrates a significantly lower percentage of task timeouts during offloading compared to other algorithms. The lower task timeout percentage indicates its effectiveness in maintaining high task completion rates, a critical metric for service quality in MEC systems. However, the plot in Fig. 8 resembles a vertically flipped version of Fig. 5, indicating that time delay plays a dominant role in the overall system cost. This suggests that the weight coefficient $\rho_1 = \rho_2 = 0.5$ does not effectively balance delay and energy consumption.

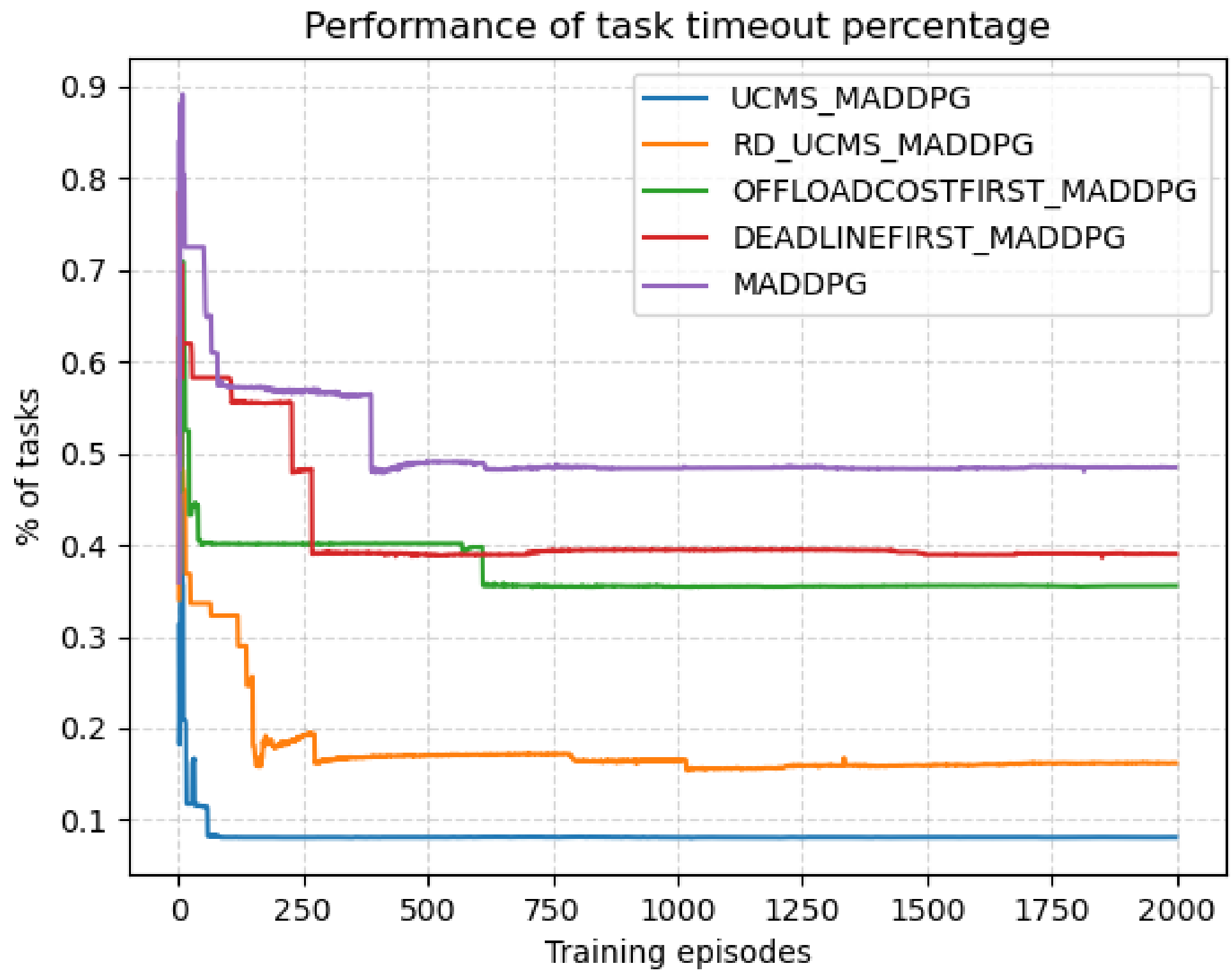

**Fig.8 Percentage of task timeouts for different algorithms**

To further assess the impact of energy consumption, we set the maximum battery threshold $b_{max}^h$ to 1MJ to increase the energy consumption penalty. The weight coefficients $\rho_1 = 1$ and $\rho_2 = 5$ are adjusted to amplify the importance of energy consumption, simulating the scenario sensitive to energy consumption. Additionally, the time slot of each episode is extended to 100 to improve the test performance. Fig. 9 shows the performance comparison of different algorithms with 100 time slots per episode. It can be observed that UCMS_MADDPG still achieves optimal performance. In contrast to the experiment with 10 time slots, MADDPG and the heuristic algorithms perform closer to UCMS_MADDPG with 100 time slots. This improvement is due to the increased number of time slots, which generates more empirical data, allowing the algorithm to generalize better.

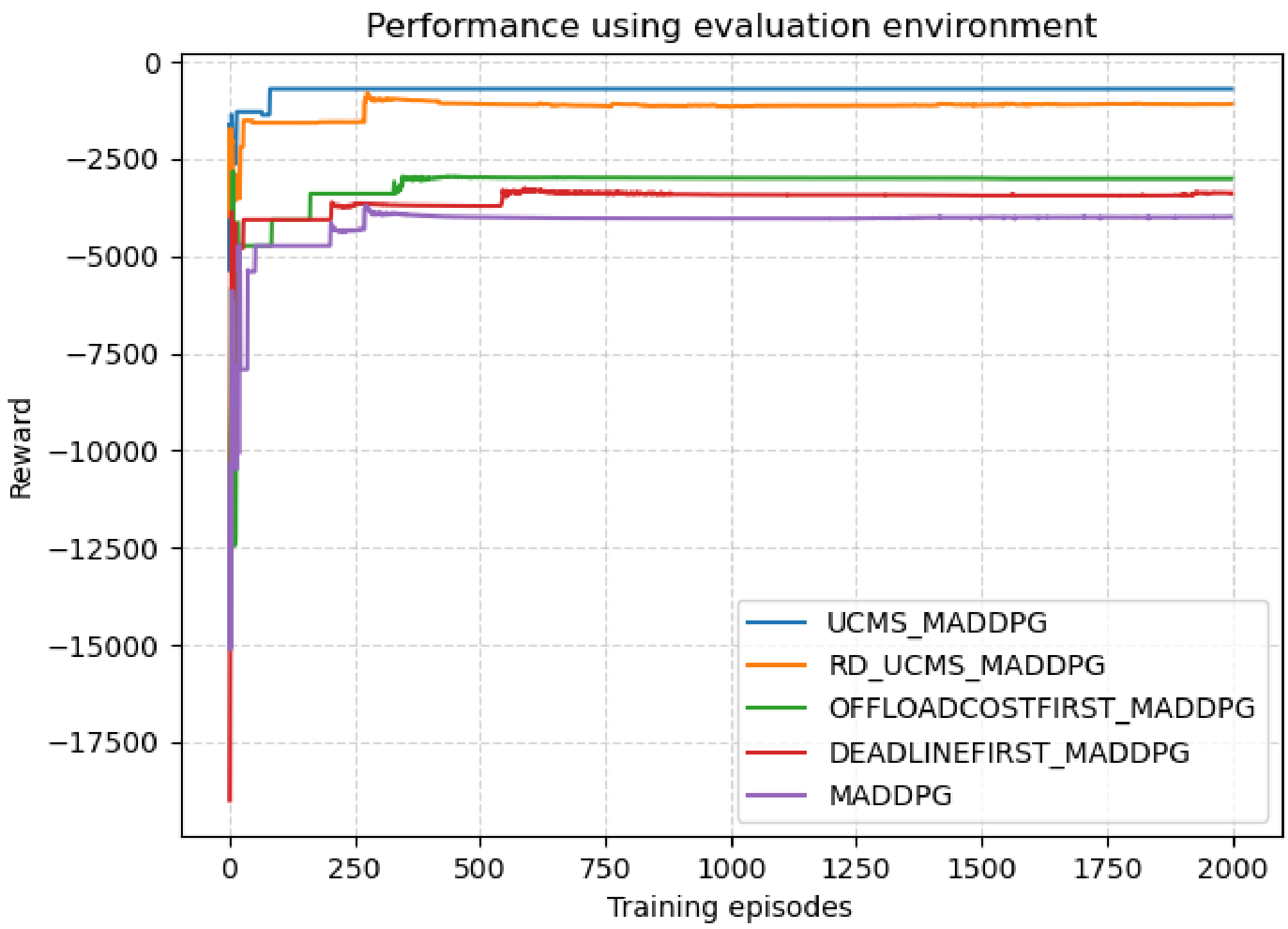

**Fig.9 Performance comparison for different algorithms in 100 time slots**

Similarly, Fig.10 illustrates the task timeout percentages for different algorithms with 100 time slots. There are significant differences between Fig. 10 and Fig. 9, indicating that modified experimental parameters have an effect. Meanwhile, Fig. 8 and Fig. 10 show that the significantly lower task timeout percentage achieved by UCMS_MADDPG reflects its effectiveness in balancing resource efficiency and service quality. By explicitly considering server storage constraints and dynamically adjusting offloading decisions based on real-time conditions, our algorithm prevents resource overcommitment that causes task drops in other approaches. Next, we examine the impact of energy consumption by comparing how frequently UCMS_MADDPG and the other four benchmark algorithms exceed the battery threshold. Fig.11 shows the percentage of UDs operating below the battery threshold for each algorithm. UCMS_MADDPG performs similarly to the other baseline algorithms in this respect. This is because the overall system return considers both latency and energy consumption, while Fig. 11 only reflects the number of UDs running below the battery threshold. According to the other experimental results, it can be concluded that UCMS_MADDPG has better overall performance and greater advantages in dynamic environments.

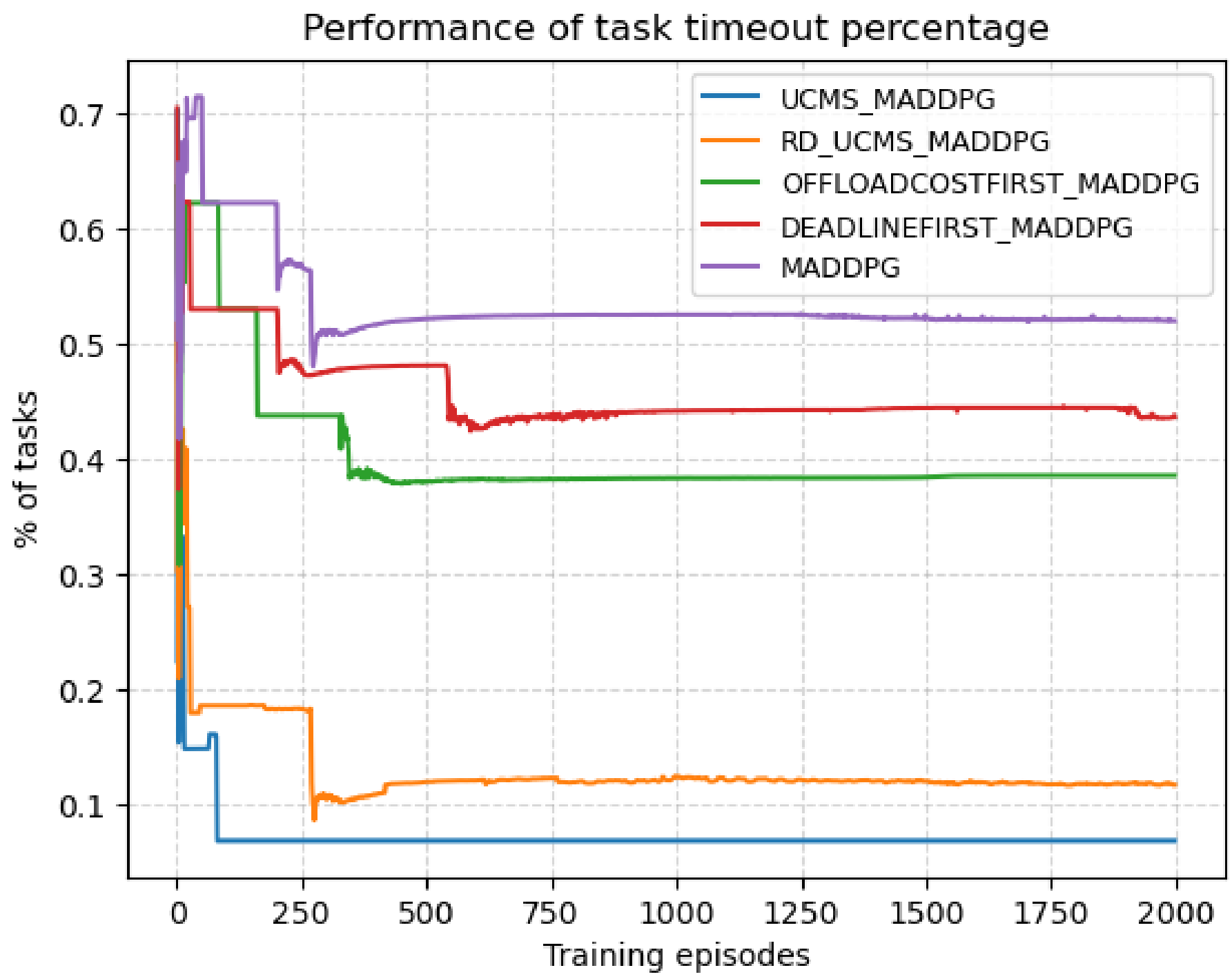

**Fig.10 Percentage of task timeouts for different algorithms in 100 time slots**

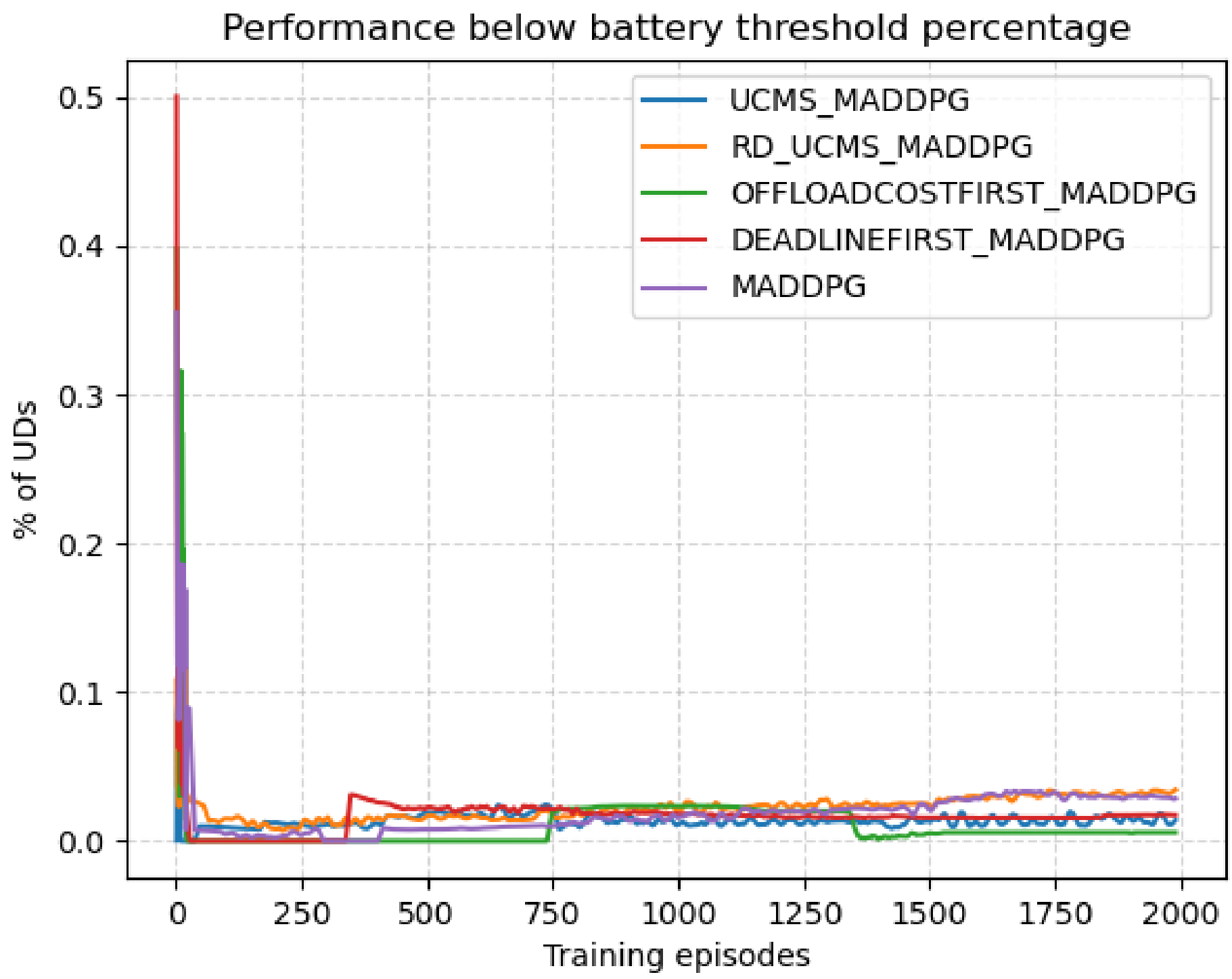

**Fig.11 Percentage below the battery threshold for different algorithms in 100 time slots**

## 5. 4 Supplementary Scalability Analysis

To further evaluate the scalability of the proposed UCMS_MADDPG framework, we extend the experiments from the baseline scenario of 3 ESs (48 UDs) to larger scenarios with 4 ESs (64 UDs) and 5 ESs (80 UDs), setting 500 training episodes while maintaining a consistent user density of 16 UDs per ES. Figs. 12 and 13 present the reward convergence curves for the training and evaluation environments, respectively, across different system scales.

In the training environment (Fig. 12), the 3 ESs configuration exhibits the fastest convergence, approaching a stable reward within approximately the first 100 episodes. In contrast, the 4 ESs and 5 ESs configurations require approximately 200-300 episodes to reach near-stability, reflecting the increased coordination complexity in larger-scale systems. Moreover, although the 4 ESs and 5 ESs curves exhibit greater fluctuations during early training, the variance gradually decreases, and all three configurations converge to relatively stable reward levels by the end of 500 episodes.

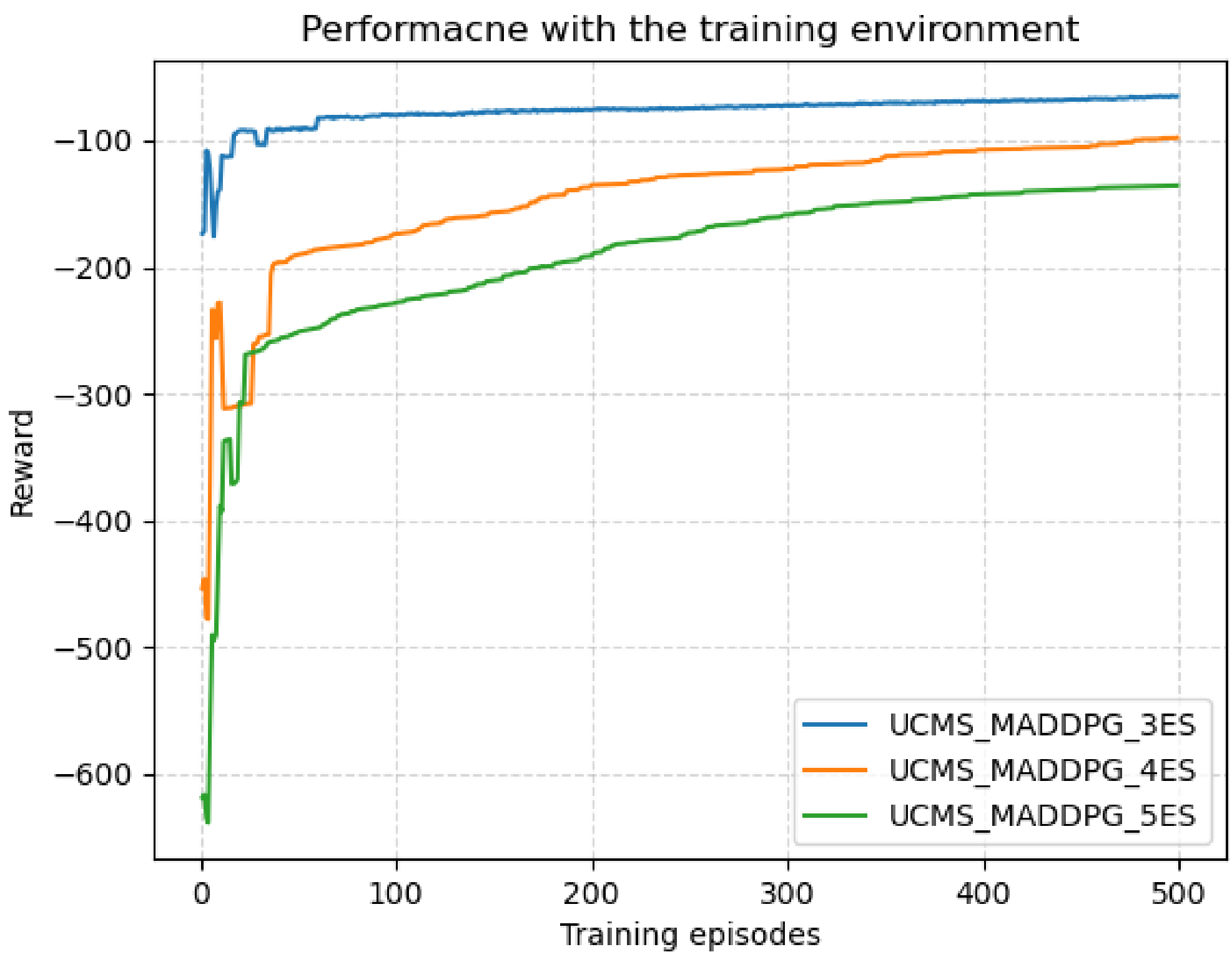

**Fig.12 Comparison of convergence performance with different numbers of ESs in the training environment**

In the evaluation environment (Fig. 13), the reward curves for 4 ESs and 5 ESs stabilize smoothly after initial oscillations, although their steady-state rewards are slightly lower than those of the 3 ESs. This observation suggests that the proposed framework can maintain robust learning capability across different system scales, despite the additional optimization challenges introduced by the increased number of servers and users. Overall, the supplementary experimental results further confirm the stable convergence performance of UCMS_MADDPG in MEC networks, demonstrating its potential scalability for larger practical deployments.

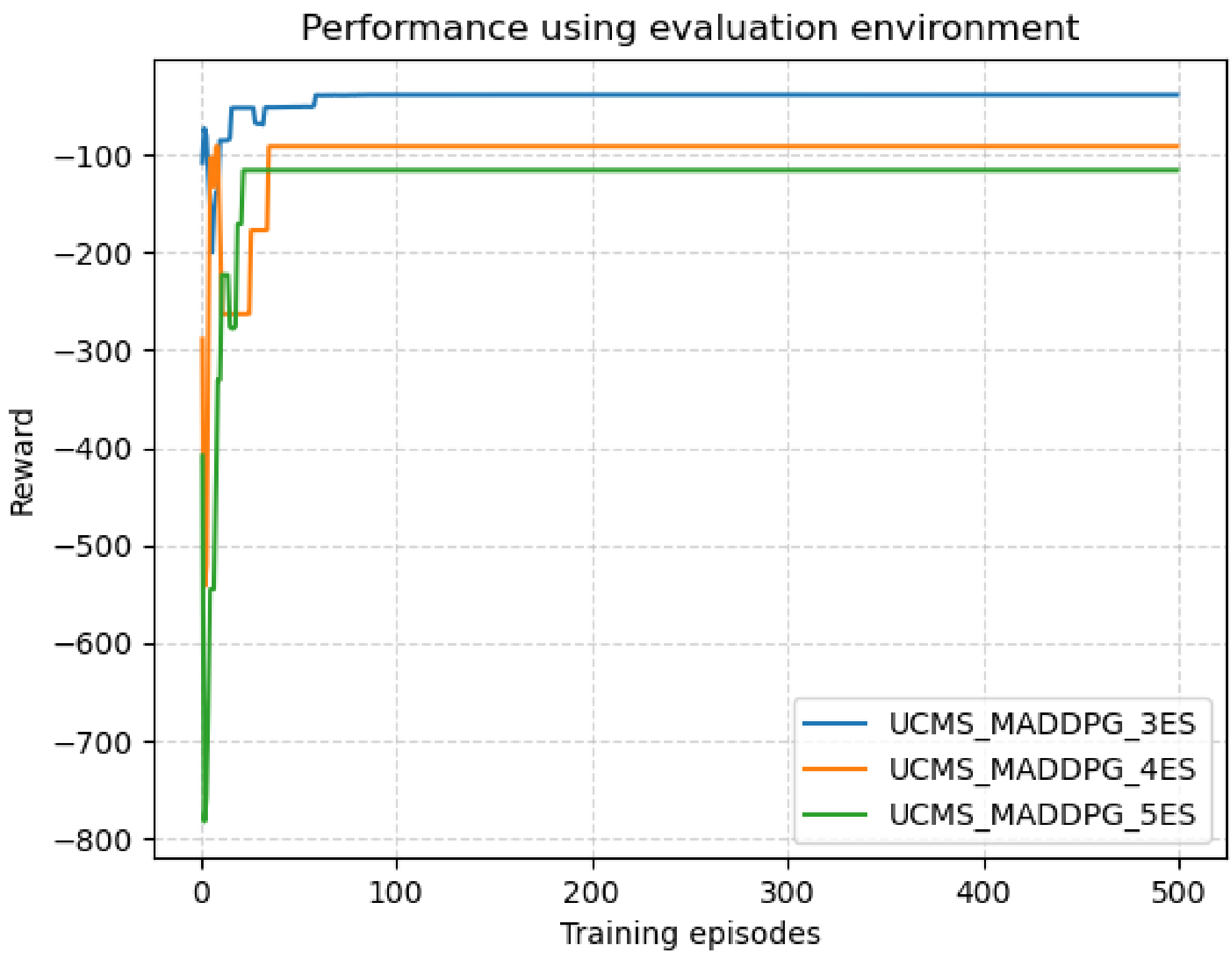

**Fig.13 Performance comparison with different numbers of ESs in the evaluation environment**

# 6 Conclusion

This paper presents a user-centric DRL model splitting inference scheme to address the challenges in dynamic MEC environments, characterized by multi-user, multi-server, and multi-angle resource constraints. The proposed scheme effectively optimizes task execution delay, energy consumption, and service quality, as evidenced by its ability to maintain low task drop rates while minimizing resource costs. We address this NP-hard non-convex MIP problem by decoupling it into tractable subproblems. Specifically, we introduce a user-server co-selection algorithm to handle the association between users and servers. Additionally, we leverage user-centric model splitting inference to design a UCMS_MADDPG-based offloading algorithm for task decision-making and response coordination. In the MDP transformation, we split the action space into user-side and server-side components to support hybrid decision-making, where the final offloading decision considers both continuous and discrete actions. Furthermore, we introduce a preferential sampling mechanism based on a reward-error trade-off to improve learning efficiency and stability. Simulation results confirm the convergence of UCMS_MADDPG and demonstrate its superior performance in dynamic environments, validating its effectiveness in optimizing delay and energy consumption and highlighting its scalability for larger practical deployments.